\documentclass[referee]{raa}            
\usepackage{lscape}
\usepackage{longtable}
\usepackage{booktabs}
\usepackage{graphicx}             
\usepackage{natbib}
\usepackage{amssymb,amsmath}
\usepackage{siunitx}
\usepackage[pagebackref=true]{hyperref}

\begin{document}
	
	\title{Selecting Optimal Stellar Calibration Fields for the CSST Imaging Survey}
	\volnopage{Vol.0 (20xx) No.0, 000--000}      
	\setcounter{page}{1}          
	
	\author{Chenxiaoji Ling
		\inst{1}
		\and Juanjuan Ren
		\inst{1}
		\and Li Shao
		\inst{1}
		\and Zhimin Zhou
		\inst{1}
		\and Peng Wei
		\inst{1}
		\and Youhua Xu
		\inst{1}
		\and Jinyu Hu
		\inst{1}
		\and Xin Zhang
		\inst{1}
		\and Su Yao
		\inst{1}
	    \and Hu Zhan
	    \inst{1}
	    \and Chao Liu
	    \inst{1}
	}
	
	\institute{National Astronomical Observatories (NAOC), Chinese Academy of Sciences (CAS), Beijing 100101, China}

	{\small Received 20xx month day; accepted 20xx month day}

    \abstract{
    The Chinese Space Station Survey Telescope (CSST) will perform a decade-long high-precision wide-field imaging survey that relies on rigorous on-orbit calibration. This necessitates stable celestial benchmark fields to maintain photometric and astrometric consistency throughout the mission lifetime. We establish comprehensive selection criteria including observational visibility, stellar number density, bright-star contamination, and interstellar dust extinction. Using the CSST Observation Strategy Analysis Tool (\texttt{COSAT}) and all-sky dust maps from \textit{Planck} and SFD, we constrain eligible regions to the ranges of ecliptic latitude $\lvert \beta \rvert \geq \SI{50}{\degree}$ and galactic latitude $\lvert b \rvert \geq \SI{15}{\degree}$. From an initial sample of 29 candidate clusters meeting these spatial constraints, six globular clusters (M13, M92, NGC 104, NGC 362, NGC 1261, and NGC 1851) are identified as optimal calibration  fields, fulfilling all the critical criteria. These selected clusters are recommended as optimal calibration field candidates for CSST’s on-orbit calibration program, and are fundamental to achieve unprecedented photometric precision in CSST's space-based survey.
    }	
	
	\authorrunning{Ling C.}            
	\titlerunning{Calibration Fields for CSST Wide-Field Survey}  
	
	\maketitle

	\section{Introduction}
	\label{sect:intro}
	
    The Chinese Space Station Survey Telescope (CSST) is a 2-meter-aperture wide-field survey telescope designed to co-orbit with the China Space Station (CSS, or Tiangong). As a Stage-IV space survey facility, it offers distinctive advantages for conducting large-field celestial surveys \citep{Gong2025}. Adopting a three-mirror optical design, the CSST achieves both a wide field of view (FOV) and high spatial resolution: across its 1.1 $\rm deg^2$ primary imaging FOV, the 80\% encircled energy radius (EER) of the point spread function (PSF) is specified to be no greater than $\SI{0.15}{\arcsecond}$. The CSST will execute a decade-long legacy survey covering 17,500 $\rm deg^2$ at mid-to-high Galactic latitudes, reaching a limiting depth of 25--26 AB mag across seven photometric bands (\textit{NUV}, $u$, $g$, $r$, $i$, $z$, $y$) that span the wavelength range of $\SIrange{255}{1000}{\nano\meter}$ \citep{zhan2021}. Operating on a $\ang{42}$ inclination orbit adjacent to the CSS, the CSST is equipped with five scientific instruments; among these, the Main Survey Camera (MSC) and the Multi-Channel Imager \citep[MCI;][]{Zheng2025} are capable of conducting high-precision multi-band imaging observations.
    
    Such high-precision multi-band observations rely on regular on-orbit calibration to monitor and correct instrumental systematics and temporal drifts. Carefully selected calibration fields serve as stable celestial benchmarks for characterizing long‑term performance and ensuring photometric and astrometric homogeneity. These fields must satisfy a suite of stringent and interlinked criteria: (1) persistent visibility for scheduled calibration observations throughout the mission lifetime; (2) appropriate stellar density to supply a  statistically robust set of reference stars; (3) minimal contamination from bright stars (e.g., $V \leq 7$~mag) to avoid detector saturation and complex stray-light effects such as ghosts, halos and reflections; (4) regions of negligible interstellar extinction to ensure the measured photometric properties reliably represent the intrinsic fluxes of the reference stars, thereby securing the fundamental photometric accuracy of the entire survey.
    
    Calibration fields selection is critically determined by each space telescope's unique scientific objectives, instrumental capabilities, and operational constraints. For example, the Hubble Space Telescope (\textit{HST}) has utilized globular clusters such as NGC 104 and NGC 2419 for calibrating the Advanced Camera for Surveys (ACS) \citep{sirianni2005}, \textit{Euclid} has designated a dedicated calibration region near the North Ecliptic Pole (NEP) within perennial visibility \citep{Euclid2022_v1}, and the \textit{Roman} Space Telescope plans to rely primarily on a field in the Large Magellanic Cloud (LMC)\footnote{\href{https://roman.gsfc.nasa.gov/science/calibration/WFI\_Touchstone\_Fields\_RevA\_2025-03-31.pdf}{WFI Touchstone Fields RevA 2025-03-31.pdf}}. However, CSST’s survey-oriented design affords it a far larger FOV than the imaging instruments of \textit{HST}, rendering the calibration fields of \textit{HST} in need of a comprehensive re-evaluation. Although CSST shares comparable survey parameters with \textit{Euclid}, the two telescopes differ fundamentally in both observational strategy and celestial visibility characteristics: stationed at the Sun–Earth L2 Lagrange point, \textit{Euclid} achieves perennial visibility of the sky around the NEP, and it can implement a contiguous tile-by-tile scanning mode for observations. By contrast, CSST orbits the Earth with inherently discontinuous observational windows, which not only renders contiguous tile-by-tile scanning unfeasible but also means no sky region across the celestial sphere is permanently observable for the telescope. Furthermore, CSST operates in the near-ultraviolet to optical wavelengths and is therefore highly sensitive to dust extinction. This makes low-extinction extragalactic fields more suitable than dust-rich regions such as the LMC, a key difference from the calibration approach of \textit{Roman}.
    
    Calibration field selection for the CSST demands a systematic approach that accounts for its distinctive observational characteristics while balancing multiple calibration requirements. In practice, star clusters represent prime candidate fields, as their intrinsic stellar density can inherently fulfill the telescope’s requirement for sufficiently populated reference sources. We therefore implement a two-stage selection: first, we apply spatial constraints to identify celestial regions meeting basic observational feasibility criteria, thereby refining the candidate cluster sample; second, within these qualified regions, we evaluate individual clusters against specific calibration constraints, including observational visibility, stellar number density, dust extinction, and bright star contamination.
    
    The paper is organized as follows. We begin by establishing spatial constraints through orbital simulations and all-sky extinction maps, defining viable regions for calibration field selection (Section \ref{s2}). Second, within these constrained regions, we perform a progressive screening of star cluster samples based on the calibration requirements (Section \ref{s3}). Section \ref{s4} presents a multi-dimensional characterization of the final candidate fields, evaluating their suitability across key parameters. Finally, we summarize our work and outline future prospects (Section \ref{s5}).

	\section{Sky Constraints for Field Selection}
	\label{s2}
	
	Directly investigating thousands of star cluster fields across the entire celestial sphere is highly inefficient. In practice, the sky distribution of star clusters exhibits strong correlations with two key selection parameters: observational visibility and interstellar extinction. Therefore, we adopt a strategy of imposing spatial constraints as a priority to efficiently refine the candidate sample. Observational visibility is primarily correlated with ecliptic latitude ($\beta$): clusters at higher $\lvert \beta \rvert$ values are significantly more accessible for observations. In contrast, interstellar dust extinction is strongly regulated by galactic latitude ($\lvert b \rvert$), and clusters located in regions with high $\lvert b \rvert$ suffer from far weaker extinction effects. This section provides the quantitative basis for these spatial constraints. We determine specific thresholds for ecliptic and galactic latitudes using systematic simulations and all-sky extinction maps.

	\subsection{Visibility Constraint from Orbital Simulations}

	For the selection of high-visibility calibration fields, we first calculate the observational windows across the entire celestial sphere based on CSST’s orbital parameters and predefined constraints. The observational visibility of CSST is determined by a combination of observational requirements and operational scheduling constraints, including Solar/Lunar/Earth avoidance angles, FOV collocation, telescope maneuvering and stabilization time, and routine maintenance operations \citep{fu2023}. An accurate assessment of such visibility therefore requires detailed simulations of CSST’s orbit and its associated observational constraints.

    In this work, we employ the CSST Observational Strategy Analysis Tool (\texttt{COSAT})\footnote{\href{https://gitee.com/cosmo-xyh/cosat}{COSAT: https://gitee.com/cosmo-xyh/cosat}} to perform high-precision visibility predictions for CSST. \texttt{COSAT} is a Python-based tool developed primarily for statistical analysis of CSST observational scheduling simulations. Its core functionalities include evaluating sky coverage and observation cadence for predefined survey sequences, as well as conducting simplified visibility analyses for arbitrary sky regions. Although \texttt{COSAT} currently omits certain engineering constraints (e.g., the energy balance and Control Moment Gyroscope thermal constraints), it remains adequate for identifying suitable calibration fields in this study.

	\subsubsection{COSAT Simulations for Observational Visibility}

    The MSC and MCI are designed for distinct observational missions, leading to significant differences in divergent requirements for their conditions. During nominal operations, the MSC and MCI are allocated to different orbital based on $\hat{\beta}$ \footnote{In this paper, the symbol $\beta$ denotes the ecliptic latitude.}, defined as the angle between the telescope's orbital plane and the direction to the Sun. The $\hat{\beta}$ angle is directly correlated with the duration of continuous observation windows: a larger $\hat{\beta}$ angle generally corresponds to longer uninterrupted observing intervals in each orbit, and vice versa. As a wide-field survey instrument, MSC executes the full-sky imaging survey following a predefined scanning schedule, and thus requires a higher $\hat{\beta}$ angle to maintain sufficiently long, continuous observational windows. In contrast, MCI is designed for targeted observations of specific celestial sources, and has less stringent requirements on continuous observing duration. To ensure the MSC receives 70\% of the total observational time \citep{zhan2021}, orbits are partitioned according to the $\hat{\beta}$ angle: those with $\hat{\beta} \geq \ang{11.6}$ are allocated to the MSC, while orbits with $\hat{\beta} < \ang{11.6}$  are assigned to other scientific instruments, including the MCI. It is important to emphasize that this simulation prioritizes the observational visibility of the MSC and MCI rather than their actual allocated observational time.

	Our \texttt{COSAT} simulations employ the latest orbit files (Version 2025) that cover the period from 2027 to 2038. When calculating the observational visibility windows for input celestial coordinates across each orbital cycle (for both the MSC and MCI), the constraints adopted in this work include the Solar/Lunar/Earth avoidance angles and the exclusion of regions within the South Atlantic Anomaly (SAA). The detailed parameter settings of these constraints are summarized in Table \ref{table:orbit}. We perform post-processing and statistical analysis of the simulation outputs using the \texttt{AstroPy} package \citep{astropy:2013, astropy:2018, astropy:2022}.
	
	\begin{table}[htbp]
	    \centering
	    \begin{tabular}{cc}
        \toprule
        \textbf{Constraints} & \textbf{Degrees (\si{\degree})}  \\ \midrule 
        Sun-avoidance & $\geq$ 50 \\
        Moon-avoidance & $\geq$ 40 \\
        Earth-avoidance (dayside) & $\geq$ 70 \\
        Earth-avoidance (nightside) & $\geq$ 30 \\
        SAA-avoidance & {--}\\ \bottomrule
        \end{tabular}
	    \caption{Orbital constraints applied in \texttt{COSAT} simulations for calculating visibility time windows.}
	    \label{table:orbit}
	\end{table}

    \subsubsection{Constraints Based on All-Sky Observational Visibility}
    \label{s21}
    
    To comprehensively characterize the observational visibility across the entire celestial sphere, we adopt the HEALPix tessellation method \citep{gorski2005} in conjunction with the \texttt{healpy} package \citep{zonca2019}, which divides the sky into 49,152 discrete regions of approximately 1 $\rm deg^2$ each. We compute the visibility of each HEALPix region using its central celestial coordinates via  \texttt{COSAT}  over the 11-year mission period from 2027 to 2038. The accuracy of this simulation is not guaranteed for periods extending beyond 2038.

    \begin{landscape}
	   	\begin{figure}
		\centering
		\includegraphics[width=1.4\textwidth]{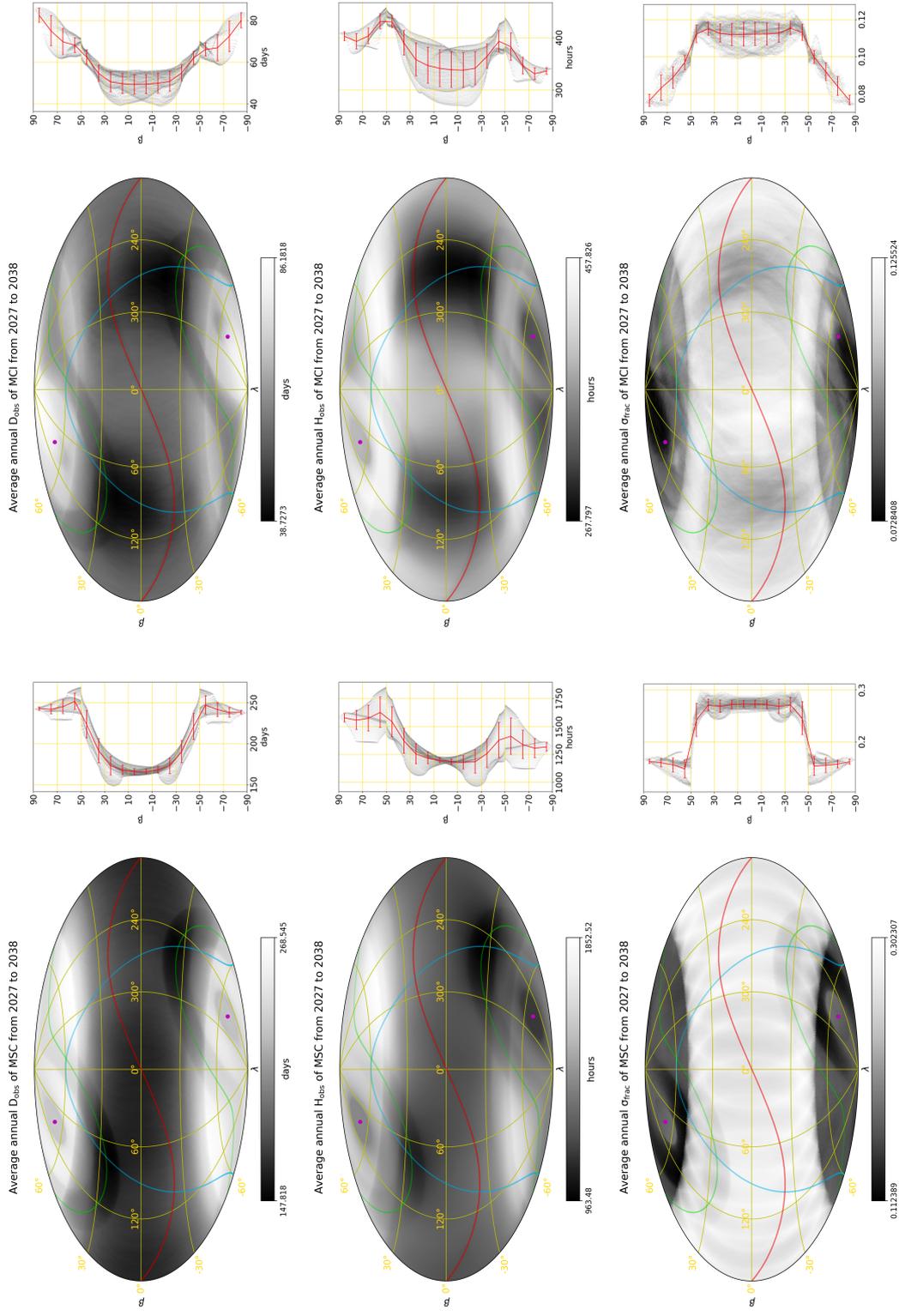}
		\caption{All-sky distribution of visibility parameters ($D_{\rm obs}$, $H_{\rm obs}$, $\sigma_{\rm frac}$) for the MSC and MCI in ecliptic coordinates from 2027 to 2038, respectively. The red and cyan lines represent the Equatorial and Galactic Planes, respectively. The magenta points represent the celestial poles and green lines represent declination at $\rm \delta = \pm \ang{48}$. Right insets show binned averages of each parameter versus ecliptic latitude. Black dots represent parameters at individual HEALPix regions; red broken lines and error bars represent binned averages ($\Delta \beta= \ang{10}$) with $1\sigma$ uncertainties. }
		\label{fig:visibility}
	    \end{figure} 
	\end{landscape}    
    
    The CSST completes dozens of orbits around the Earth each day. For a given sky region, there exist both observable and unobservable segments within each orbit. Given the computational complexity of simulations, we establish a set of calibration-specific observational criteria to quantify observational visibility: (1) Observable orbits are defined as orbital segments permitting continuous observations with $T_{\rm orbit} \geq \SI{5}{\minute}$.  This threshold accommodates the nominal exposure time (\SI{150}{\second} for Wide-field Survey) per calibration visit, including  settling and readout overheads. The orbits with duration shorter than this threshold are excluded as they are insufficient to complete a single standard calibration exposure; (2) Observable days are defined as dates on which the cumulative observation time across all valid observable orbits satisfies $\sum T_{\rm orbit} \geq \SI{90}{\minute}$, thus providing sufficient time for multiple calibration exposures in a single day; (3) Observable hours are defined as the cumulative observation time (in hours) across all qualified observable orbits for a given sky region.
    
    Based on the above three calibration criteria, we define three annual statistical parameters to quantify the long-term visibility characteristics from the simulation results:
    \begin{itemize}
        \item $D_{\rm obs}$: the annual count of observable days for a given sky region.
        \item $H_{\rm obs}$: the total annual observable hours accumulated across all observable days for a given sky region.
        \item $\sigma_{\rm frac}$: the standard deviation of the monthly fractions $frac$ over twelve months in a year, where $frac$ is defined as the ratio of the number of observable days to the total number of days in a given month.
    \end{itemize}
    
    $D_{\rm obs}$ and $H_{\rm obs}$ jointly quantify the overall annual observational visibility of a sky region, while $\sigma_{\rm frac}$ characterizes the monthly variability in the number of observable days. We compute the values of these three statistical parameters for each HEALPix region annually, and further derive their average values over the entire 11-year study period. Figure \ref{fig:visibility} presents the all-sky distributions of these parameters for both MSC and MCI, while Table \ref{tab:latitude_stats} provides their mean values binned by ecliptic latitudes.

    All three statistical parameters demonstrate a strong dependence on ecliptic latitude, with $\lvert \beta \rvert = \ang{50}$ emerging as a critical threshold. For the MSC, the  average annual of $D_{\rm obs}$ reaches $\sim 240$ days in high-latitude regions ($\lvert \beta \rvert \geq \ang{50}$), decreasing to $\sim 170$ days in low-latitude regions ($\lvert \beta \rvert \leq \ang{30}$). Significant variability occurs within the transitional latitude range ($\ang{30} \leq \lvert \beta \rvert \leq \ang{50}$), accompanied by a slight reduction ($\sim 5\%$) in observable days near the celestial poles. The average annual $\sigma_{\rm frac}$ increases sharply when  $\lvert \beta \rvert $ decreases to around $\ang{50}$, indicating that the monthly distribution of observable days is more uniform at high ecliptic latitudes than at low ecliptic latitudes. Regions with $\lvert \beta \rvert \leq \ang{50}$ experience 2--3 consecutive completely unobservable months for calibration observations. The average annual of $H_{\rm obs}$ exhibits hemispheric asymmetry resulting from SAA avoidance constraints: northern high-latitude regions ($\beta \geq \ang{50}$) yields  $\sim 1600$ hours annually, which is approximately $20\%$ higher than those of southern high-latitude regions ($\beta \leq -\ang{50}$; $\sim 1300$ hours). In contrast, near-ecliptic regions ($\lvert \beta \rvert \leq \ang{20}$) provide $\sim 1200$ observable hours annually, about $30\%$ lower than those of northern high-latitude regions. The visibility trends for MCI are consistent with those for MSC, but exhibit larger statistical variance. These findings further reinforce the need for prioritizing regions with high ecliptic latitudes ($\lvert \beta \rvert \geq \ang{50}$) to ensure both stable and sufficient observational visibility.

    We further note that visibility is not only correlated with ecliptic latitude, but also exhibits measurable dependencies on declination ($\delta$) and ecliptic longitude ($\lambda$) (see Figure \ref{fig:visibility}). For both the MSC and MCI, visibility decreases near $\delta \sim \pm\ang{90}$ (the celestial poles) and $\delta \sim \pm \ang{48}$ (perpendicular to the telescope’s orbital plane). Earth occultation renders regions near the celestial poles unobservable when the telescope orbits above the opposite hemisphere, reducing observable hours around both poles. Similarly, regions along the celestial axis perpendicular to the telescope’s orbital plane also become inaccessible due to Earth occultation. These regions gradually become observable again following orbital precession over time, leading to a systematic reduction in observable time for the entire toroidal region centered at $\delta \sim \pm\ang{48}$. MCI additionally shows two distinct regions of reduced visibility along the ecliptic longitude, arising from the resonant effect between its observation windows and lunar orbital position, which suppresses visibility at specific ecliptic longitudes.

    \begin{table}[htbp]
      \centering
      \small  
      \begin{tabular}{ccccccc}
        \toprule  
        \textbf{$\si{\beta}$ ($\si{\degree}$)} & \multicolumn{3}{c}{\textbf{MSC}} & \multicolumn{3}{c}{\textbf{MCI}} \\
        \cmidrule(lr){2-4} \cmidrule(lr){5-7} %
         &$\overline{D_{\rm obs}}$ & $\overline{H_{\rm obs}}$ & $\overline{\sigma_{\rm frac}}$ & $\overline{D_{\rm obs}}$ & $\overline{H_{\rm obs}}$ & $\overline{\sigma_{\rm frac}}$ \\  
        \midrule  
        \SIrange{70}{90}{\degree}  & 242 & 1564.4 & 0.160 & 77 & 395.7 & 0.082 \\
        \SIrange{50}{70}{\degree}   & 249 & 1606.4 & 0.151 & 69 & 419.7 & 0.094 \\
        \SIrange{30}{50}{\degree}   & 206 & 1452.5 & 0.258 & 58 & 408.8 & 0.114 \\
        \SIrange{10}{30}{\degree}   & 171 & 1236.6 & 0.271 & 50 & 352.5 & 0.112 \\
        \SIrange{-10}{10}{\degree}  & 166 & 1192.1 & 0.274 & 49 & 340.2 & 0.112 \\
        \SIrange{-30}{-10}{\degree} & 171 & 1183.1 & 0.271 & 50 & 339.8 & 0.113 \\
        \SIrange{-50}{-30}{\degree} & 205 & 1315.9 & 0.260 & 58 & 378.0 & 0.114 \\
        \SIrange{-70}{-50}{\degree} & 245 & 1383.4 & 0.154 & 67 & 367.6 & 0.097 \\
        \SIrange{-90}{-70}{\degree}  & 238 & 1310.9 & 0.158 & 74 & 332.5 & 0.082 \\
        \bottomrule  
      \end{tabular}
      \caption{The average annual of $D_{\rm obs}$ (days), $H_{\rm obs}$ (hours) and $\sigma_{\rm frac}$ of MSC and MCI in different ecliptic latitude bins.} 
      \label{tab:latitude_stats} 
    \end{table}

    Among the three statistical parameters, monthly consistency of observable days is adopted as the primary selection criterion, reflecting the priority placed on stable, month-to-month visibility of calibration fields. Extended periods of one to two months with complete non-visibility are deemed unacceptable for calibration field selection. Consequently, we impose a primary constraint on ecliptic latitude, requiring $\lvert \beta \rvert \geq \ang{50}$ for suitable calibration regions.
    
    \subsection{Extinction Constraints from Dust Maps}
    \label{s2.2}
    
    \begin{figure}[htbp]
        \centering
        \includegraphics[width=\textwidth]{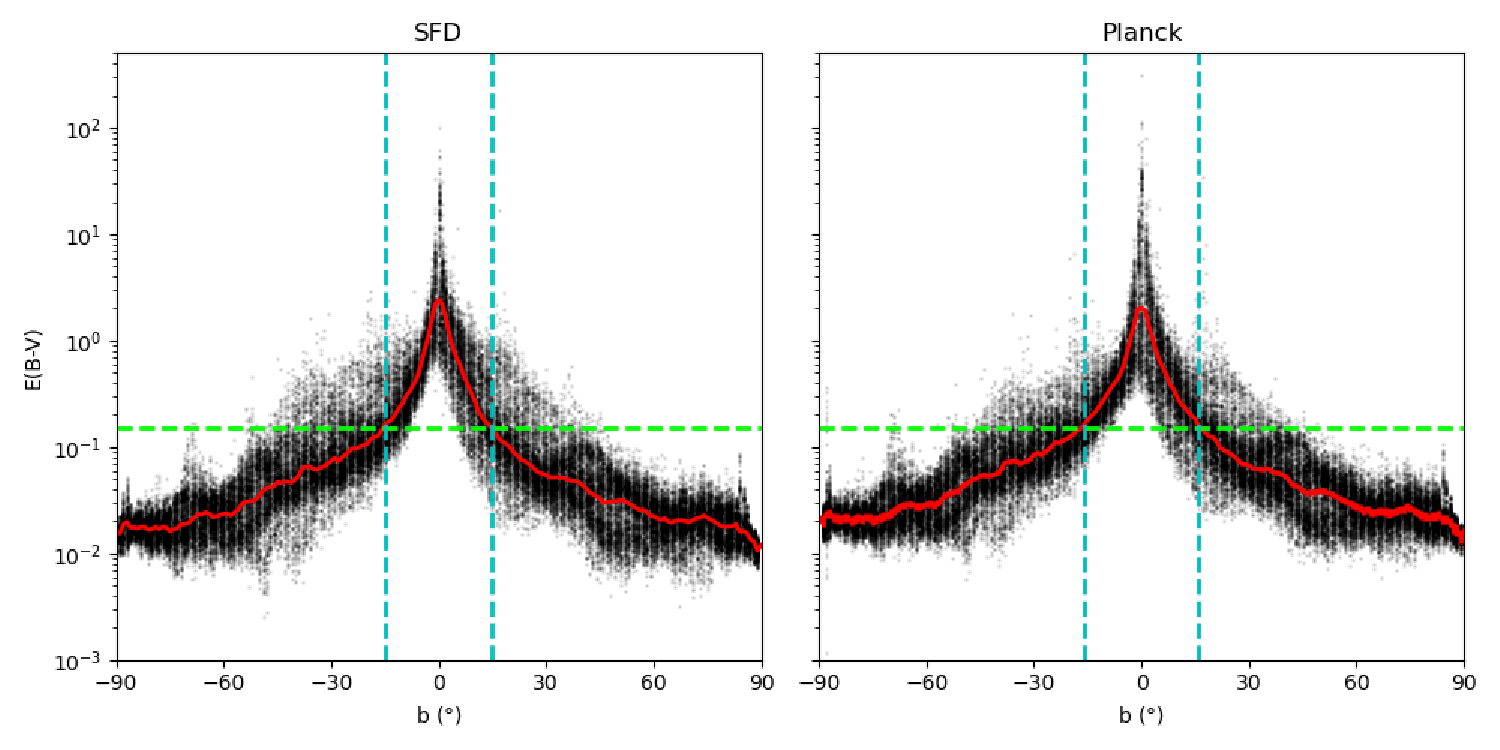}
        \caption{The distribution of $E(B-V)$ as a function of Galactic Latitude ($b$). The black points represent the $E(B-V)$ values measured at the systematic grid sampling points. The red solid line denotes the median $E(B-V)$ value at each galactic latitude. The green dashed line corresponds to the extinction threshold $E(B-V) = 0.15$. The cyan dashed lines mark the intersections between the red median line and the green threshold line. These intersections are located at $b = \pm \ang{15}$ for the SFD map and $b = \pm \ang{16}$ for the \textit{Planck} map, respectively.}
        \label{fig:ebv_hist}
    \end{figure}
    
    Interstellar extinction critically affects on-orbit calibration field selection, particularly by diminishing the number of detectable targets in bluer photometric bands (e.g., the $\textit{NUV}$ and $u$ bands). Ideal calibration regions should avoid areas with significant interstellar extinction to ensure photometric accuracy. In this work, we use the \texttt{dustmap} package \citep{green2018} to combine the all-sky dust extinction maps from SFD \citep{sfd} and \textit{Planck} \citep{planck2016}, constructing a high-precision all-sky dust extinction model. This approach enables accurate calculation of the $E(B-V)$ extinction value for celestial sky positions.
    
    We perform systematic grid sampling across the celestial sphere with the following specifications:
    \begin{itemize}
        \item galactic longitude ($l$): Sampled from $\ang{0}$ to $\ang{360}$ at $\ang{1}$ intervals across the full sky.
        \item galactic latitude ($b$): Sampled from $\ang{-90}$ to $\ang{90}$ at $\ang{1}$ intervals.
    \end{itemize}
    
    In turn, we construct an all-sky grid containing 65,160 sampling points (360 grids in galactic longitude and 181 grids in galactic latitude). Dust extinction, quantified by the $E(B-V)$ value, exhibits a strong dependence on galactic latitude. Based on comprehensive analysis, we adopt $E(B-V) \leq 0.15$ as the threshold for defining low dust extinction regions \citep{Yasuda2007}. Under this criterion, low-extinction regions correspond to $\lvert b \rvert \geq \ang{15}$ for the SFD map and $\lvert b \rvert \geq \ang{16}$ for the \textit{Planck} map, respectively (see Figure \ref{fig:ebv_hist}). By synthesizing the results from both dust models, we ultimately recommend adopting a conservative constraint of $\lvert b \rvert \geq \ang{15}$, which achieves a reasonable balance between minimizing interstellar extinction and maintaining adequate sky coverage. It should be emphasized that this galactic latitude constraint is only applied to the initial screening of star cluster candidates. The detailed dust extinction conditions around each cluster still require subsequent, case-by-case evaluations.

    \section{Candidate Clusters Selection}
    \label{s3}
        
        Star clusters serve as highly promising candidate regions for on-orbit calibration due to their inherently high stellar densities, which provide abundant reference stars for calibration. We begin our selection from the Milky Way Star Clusters (MWSC) catalog \citep{kharchenko2012}, which comprises 3006 star clusters in the Milky Way identified from the 2MAst (2MASS with Astrometry) catalog \citep{Skrutskie2006}. This catalog is sufficiently comprehensive for our analysis, given our focus on identifying bright star clusters suitable for on-orbit calibration. The catalog classifies objects into seven distinct subtypes. We retain only globular and open clusters for further analysis; to this end, we carefully inspect the \texttt{Type}, \texttt{ n\_Type}, \texttt{SType} columns and remove all other object types. We further apply the coordinate selection criteria established in the previous section ($\lvert \beta \rvert \geq \ang{50}$, $\lvert b \rvert \geq \ang{15}$) to ensure high observational visibility and low dust extinction for the candidate clusters. Applying these constraints yields 10 globular clusters and 19 open clusters meeting all criteria. Their detailed physical and observational properties are summarized in Table \ref{table:cluster_details}.
        
        In addition to observational visibility and interstellar dust extinction constraints, a qualified calibration field must also exhibit an appropriate stellar number density, while minimizing bright star contamination for the field. Furthermore, we need to conduct a case-by-case inspection of dust extinction for each cluster to identify and mitigate localized anomalies.
        
        \subsection{Stellar Number Density}
        
        Stellar number density serves as the primary constraint for calibration field suitability. Based on preliminary studies by the on-orbit calibration team, candidate calibration regions are required to exhibit stellar number densities of 50--1000 $\rm stars\ arcmin^{-2}$ for Gaia $G$-band $16 \leq G \leq 22$. To quantify this metric for pre-selected clusters, we use the \texttt{astroquery} package \citep{astroquery} to retrieve Gaia Data Release 3 (DR3) stellar data \citep{Gaia2016, Gaia2023} within each cluster field. We conduct two targeted queries to characterize stellar density at different spatial scales: (1) Counting stars within a $\SI{11}{\arcminute} \times \SI{11}{\arcminute}$ region centered on each cluster, matching the CSST CCD chip, to derive the chip-scale stellar number density; (2) Counting stars within a $\ang{1} \times \ang{1}$ field (approximate the FOV of the telescope) centered on each cluster, to evaluate the average stellar number density across the telescope's FOV. All stellar number density results are summarized in Table \ref{table:cluster_details}.
        
        Ultimately, our analysis identifies six globular clusters that satisfy the aforementioned chip-scale stellar number density criterion ($\SI{11}{\arcminute} \times \SI{11}{\arcminute}$), including M13 (NGC 6205), M92 (NGC 6341), NGC 104, NGC 362, NGC 1261, and NGC 1851. At the $\ang{1} \times \ang{1}$ field-scale stellar number density, NGC 104 and M13 emerge as the optimal candidates, possessing the highest stellar number densities among all pre-selected clusters. Furthermore, these two clusters also have the largest angular diameters of all candidate globular clusters. Only one open cluster, NGC 1901, meets the stellar number density criterion for CSST calibration. We note that for the majority of globular clusters, the stellar number density in a chip-scale field is higher than that within the $\ang{1} \times \ang{1}$ FOV. This is consistent with the radial stellar number density profile of globular clusters, which exhibits a declining trend from the cluster center to the outskirts. In contrast, the difference in stellar number density between these two spatial scales is not significant for open clusters. We note a resolution-related discrepancy between Gaia-derived densities and those expected for CSST observations, making these estimates approximate.
        
        \subsection{Bright Star Avoidance}
        
        Minimizing bright star contamination is considered the second key constraint for calibration field selection. On-orbit calibration observations are significantly affected by the presence of bright stars within or near the FOV. Ground laboratory testing results indicate that fields containing stars with $V \leq 7$ within a radius of $\ang{1}$ must be strictly avoided for calibration tasks. For the larger $\ang{1.5}$ radius, the presence of stars with $V \leq 7$ demands particular caution, and such stars should be avoided whenever feasible during observational planning. Additionally, fields contaminated by nearby bright nebulae or extended galaxies are excluded from consideration. Finally, since stars with $V \leq 10$  mag are always present within a $\ang{1}$ radius, their impact must be carefully evaluated and minimized when designing observational plans.
        
        In this work, bright stars are identified from the Tycho-2 catalog \citep{hog2000}. For each candidate calibration field, we quantify bright star contamination by counting stars with $V \leq 7 $ and $V \leq 10 $ within $\ang{1}$ and $\ang{1.5}$ radii, respectively. All selected fields undergo further visual inspection using archival survey images (DSS and SDSS) to verify the absence of significant contamination from extended objects, such as bright nebulae and galaxies. These evaluation results are documented in Table \ref{table:cluster_details}.
        
        After applying the constraint that no star with $V \leq 7$ may lie within a $\ang{1}$ radius of the field center, we identify the following clusters as qualifying:
        \begin{itemize}
            \item Globular clusters: M13 (NGC 6205), M92 (NGC 6341), NGC 104, NGC 362, NGC 1261, NGC 1851, NGC 6229.
            \item Open clusters: FSR 0504, FSR 1440, FSR 1577, FSR 1626, FSR 1629, FSR 1631.
        \end{itemize}
    
        \subsection{Effects of Dust Extinction}
        
        Lastly, the detailed dust extinction properties of each cluster field should be re-evaluated to avoid potential localized anomalies. In Section \ref{s2.2}, we investigate the global impact of dust extinction on the selection of calibration fields. In this subsection, we focus specifically on the dust extinction effects for individual candidate star clusters. Two independent extinction measurements are considered for each candidate cluster: one is the intrinsic cluster extinction $E(B-V)_{\rm cl}$, obtained directly from the MWSC catalog; the other is the line-of-sight dust extinction, derived from all-sky dust extinction maps. In this analysis, we exclusively adopt the dust extinction values from the \textit{Planck} dust map.
    
        \begin{landscape}
        \setlength{\tabcolsep}{4pt}  
        \begin{longtable}{ccccccccccccc} 
            \toprule
            $\rm Name$ & $\rm Type$ & $\rm RA(\si{\degree})$ &  $\rm Dec (\si{\degree})$ & $\rm D({\rm arcmin})$ & $\rm \rho_{\rm chip}$ & $\rm \rho_{\ang{1}\times\ang{1}}$ & $\rm n_{\rm S7}(\ang{1})$ & $\rm n_{\rm S10}(\ang{1})$ & $\rm n_{\rm S7}(\ang{1.5})$ & $E(B-V)_{\rm \rm cl}$ & $E(B-V)_{\rm mean}$ & $E(B-V)_{\rm max}$ \\ \midrule
           \endfirsthead
            \toprule
            $\rm Name$ & $\rm Type$ & $\rm RA(\si{\degree})$ &  $\rm Dec (\si{\degree})$ & $\rm D({\rm arcmin})$ & $\rm \rho_{\rm chip}$ & $\rm \rho_{\ang{1}\times\ang{1}}$ & $\rm n_{\rm S7}(\ang{1})$ & $\rm n_{\rm S10}(\ang{1})$ & $\rm n_{\rm S7}(\ang{1.5})$ & $E(B-V)_{\rm \rm cl}$ & $E(B-V)_{\rm mean}$ & $E(B-V)_{\rm max}$ \\ \midrule
            \endhead   
            M13 (NGC 6205) & g & 250.422 & 36.46 & 33 & 263.025 & 13.996 & 0 & 22 & 0 & 0.021 & 0.025 & 0.037 \\   
            M92 (NGC 6341) & g & 259.281 & 43.136 & 28.8 & 159.934 & 8.001 & 0 & 13 & 1 & 0.021 & 0.026 & 0.046 \\   
            NGC 1851 & g & 78.5325 & -40.043 & 18 & 120.81 & 5.615 & 0 & 19 & 0 & 0.021 & 0.049 & 0.095 \\   
            NGC 104 & g & 6.0045 & -72.081 & 63.6 & 105.008 & 30.881 & 0 & 19 & 1 & 0.042 & 0.040 & 0.059 \\   
            NGC 362 & g & 15.825 & -70.847 & 17.4 & 98.537 & 9.784 & 0 & 20 & 2 & 0.052 & 0.054 & 0.441 \\   
            NGC 1261 & g & 48.0675 & -55.216 & 10.2 & 65.57 & 2.978 & 0 & 14 & 0 & 0.010 & 0.021 & 0.040 \\   
            NGC 1901 & o & 79.56 & -68.44 & 22.2 & 52.826 & 57.931 & 2 & 32 & 3 & 0.021 & 0.632 & 3.274 \\   
            NGC 2298 & g & 102.2475 & -36.005 & 9.6 & 41.967 & 5.936 & 1 & 28 & 2 & 0.139 & 0.187 & 0.293 \\   
            NGC 6229 & g & 251.745 & 47.528 & 9.6 & 21.463 & 1.757 & 0 & 19 & 1 & 0.010 & 0.028 & 0.045 \\   
            IC 4499 & g & 225.0765 & -82.214 & 10.2 & 14.438 & 1.964 & 1 & 22 & 3 & 0.229 & 0.197 & 0.287 \\   
            NGC 2243 & o & 97.392 & -31.285 & 10.2 & 13.058 & 4.467 & 1 & 31 & 4 & 0.062 & 0.089 & 0.234 \\   
            Stephenson 1 & o & 283.515 & 36.805 & 39 & 7.744 & 7.691 & 4 & 51 & 5 & 0.031 & 0.096 & 0.155 \\   
            FSR 1303 & o & 101.1825 & -31.893 & 7.2 & 5.727 & 5.552 & 4 & 41 & 8 & 0.260 & 0.115 & 0.176 \\   
            Alessi 3 & o & 109.05 & -46.617 & 20.4 & 4.355 & 4.539 & 3 & 52 & 3 & 0.110 & 0.220 & 0.326 \\   
            E 3 & g & 140.238 & -77.282 & 7.2 & 4.124 & 1.493 & 1 & 23 & 1 & 0.300 & 0.229 & 0.421 \\   
            NGC 2516 & o & 119.49 & -60.75 & 48.6 & 3.81 & 3.671 & 8 & 109 & 12 & 0.071 & 0.223 & 0.312 \\   
            ESO 309-03 & o & 102.663 & -42.383 & 7.2 & 3.645 & 3.525 & 2 & 33 & 2 & 0.416 & 0.159 & 0.257 \\   
            ESO 123-26 & o & 118.05 & -60.34 & 13.2 & 3.116 & 3.338 & 10 & 97 & 10 & 0.129 & 0.217 & 0.326 \\   
            FSR 1440 & o & 110.9055 & -53.947 & 12 & 3.083 & 3.355 & 0 & 23 & 3 & 0.521 & 0.199 & 0.248 \\   
            FSR 1607 & o & 173.25 & -77.825 & 10.2 & 2.653 & 2.35 & 2 & 30 & 5 & 0.104 & 0.197 & 0.546 \\   
            ESO 425-06 & o & 91.215 & -29.185 & 6.6 & 2.612 & 2.73 & 1 & 27 & 1 & 0.146 & 0.042 & 0.056 \\  
            ESO 021-06 & o & 213.966 & -78.515 & 10.8 & 2.562 & 2.498 & 3 & 39 & 3 & 0.104 & 0.173 & 0.218 \\   \midrule
            FSR 1470 & o & 110.7 & -59.199 & 12 & 2.281 & 2.359 & 1 & 29 & 4 & 0.416 & 0.210 & 0.300 \\   
            FSR 1577 & o & 142.1925 & -76.085 & 10.2 & 1.744 & 1.764 & 0 & 30 & 4 & 0.104 & 0.188 & 0.320 \\   
            FSR 1631 & o & 191.43 & -81.258 & 13.2 & 1.463 & 1.265 & 0 & 10 & 0 & 0.104 & 0.305 & 0.903 \\   
            FSR 1629 & o & 188.73 & -80.95 & 11.4 & 1.281 & 0.96 & 0 & 12 & 0 & 0.062 & 0.339 & 0.924 \\   
            FSR 1626 & o & 185.025 & -81.51 & 13.8 & 1.157 & 1.101 & 0 & 18 & 0 & 0.208 & 0.251 & 0.766 \\   
            FSR 0504 & o & 1.455 & 81.84 & 5.4 & 0.62 & 0.512 & 0 & 22 & 2 & 0.104 & 0.182 & 0.304 \\   
            NGC 188 & o & 11.85 & 85.255 & 35.4 & 0.504 & 0.316 & 1 & 36 & 2 & 0.085 & 0.100 & 0.215 \\ \bottomrule
            \caption{This table summarizes the details of the studied clusters.  The type ``g" represents globular cluster and ``o" represents open cluster in the \texttt{Type} column. $\rm D({\rm arcmin})$ represents the angular diameter of the star clusters from the \texttt{$\rm r_1$} column in the MWSC catalog. $\rm \rho_{chip}$ and $\rm \rho_{\ang{1} \times \ang{1}}$ are the stellar number densities of Gaia $G$-band stars within a single CSST CCD chip ($\SI{11}{\arcminute} \times \SI{11}{\arcminute}$) and a $\ang{1} \times \ang{1}$ field, respectively. $n_{\rm S7}$ and $n_{\rm S10}$ are the numbers of Tycho-2 stars brighter than $V=7$ and $V=10$ mag, within a search radius $r\leq \ang{1}$ and $r \leq \ang{1.5}$, respectively. $E(B-V)_{\rm cl}$ is the intrinsic extinction of each star clusters from the MWSC catalog. $E(B-V)_{\rm mean}$ and $E(B-V)_{\rm max}$ are the mean and maximum values of $E(B-V)$ from the \textit{Planck} dust extinction map within a search radius $r\leq \ang{1}$.}
            \label{table:cluster_details}
        \end{longtable}
    \end{landscape}

        Using the \texttt{dustmap} package, we perform sampling with a resolution of $\SI{1}{\arcminute}$ within a radius of $\ang{1}$ centered on each cluster, and compute the corresponding \textit{Planck} $E(B-V)$ values for all sampled positions. We then derive both the mean ($E(B-V)_{\rm mean}$) and maximum ($E(B-V)_{\rm max}$) extinction values across the sampled region for each cluster. The results are presented in Table \ref{table:cluster_details}.
        
        Guided by the discussion in Section \ref{s2.2}, we impose dual extinction constraints: $\rm E(B-V)_{cl} \leq 0.15$ and $\rm E(B-V)_{mean} \leq 0.15$ for all candidate clusters. After applying these constraints to the initial sample of 29 star clusters, the remaining qualified targets are listed below:
        \begin{itemize}
            \item Globular clusters: M13 (NGC 6205), M92 (NGC 6341), NGC 104, NGC 362, NGC 1261, NGC 1851, NGC 6229.
            \item Open clusters: NGC 188, NGC 2243, ESO 425-06, Stephenson 1.
        \end{itemize}
    	
    	\begin{figure}[htbp]
    		\centering
    		\includegraphics[width=\textwidth]{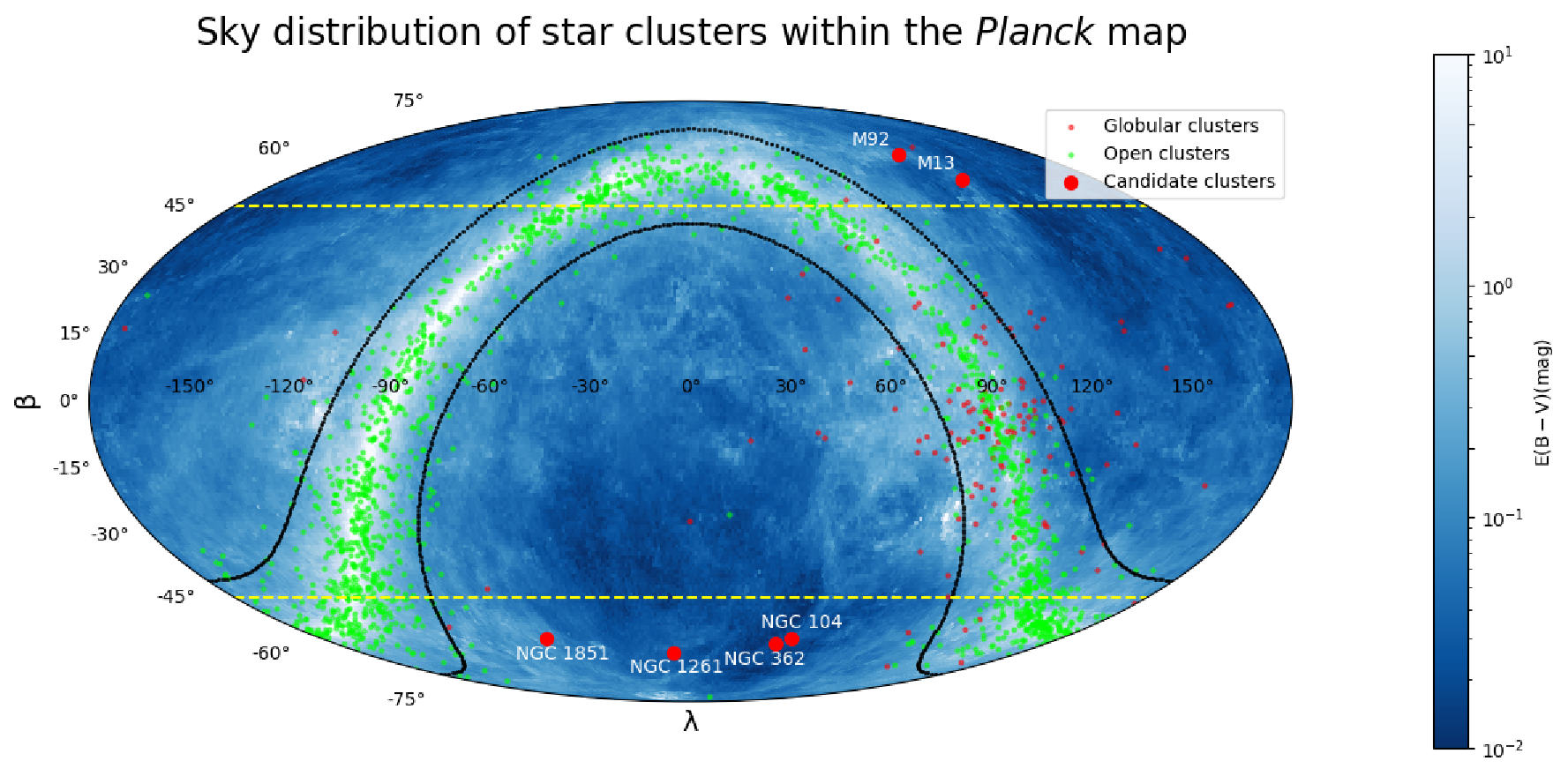}
    		\caption{All-sky distribution of candidate star clusters (green = open clusters, red = globular clusters) superimposed on the \textit{Planck} $E(B-V)$ dust extinction map with Mollweide projection in ecliptic coordinates. Dashed yellow lines indicate the ecliptic latitude constraint $\lvert \beta \rvert =\ang{50}$, while dotted black curves show the galactic latitude constraint $\lvert b \rvert =\ang{15}$. The six recommended globular clusters are highlighted with enlarged red circles.}
    		\label{fig:all_sky}
    	\end{figure}

	\section{Results and Analysis}
	\label{s4}
	
	Our comprehensive analysis in Sections \ref{s2} and \ref{s3} yields only six globular clusters meeting the core criteria for on-orbit calibration fields: M13 (NGC 6205), M92 (NGC 6341), NGC 104, NGC 362, NGC 1261, and NGC 1851. We recommend these as high-priority candidate calibration fields for subsequent in-depth investigations. However, a comprehensive assessment necessitates an understanding of their detailed performance characteristics as calibration fields. 
	
	This section provides a detailed analysis of the six candidate fields, in terms of their sky distribution, observational visibility, distribution of nearby bright stars, and spatial inhomogeneity of dust extinction. The results offer direct inputs for planning CSST calibration observations.
	
    \subsection{Sky distribution of Candidate Clusters}
    \label{s4_1}
    
\begin{figure}[htbp]
        \centering
        \includegraphics[width=\textwidth]{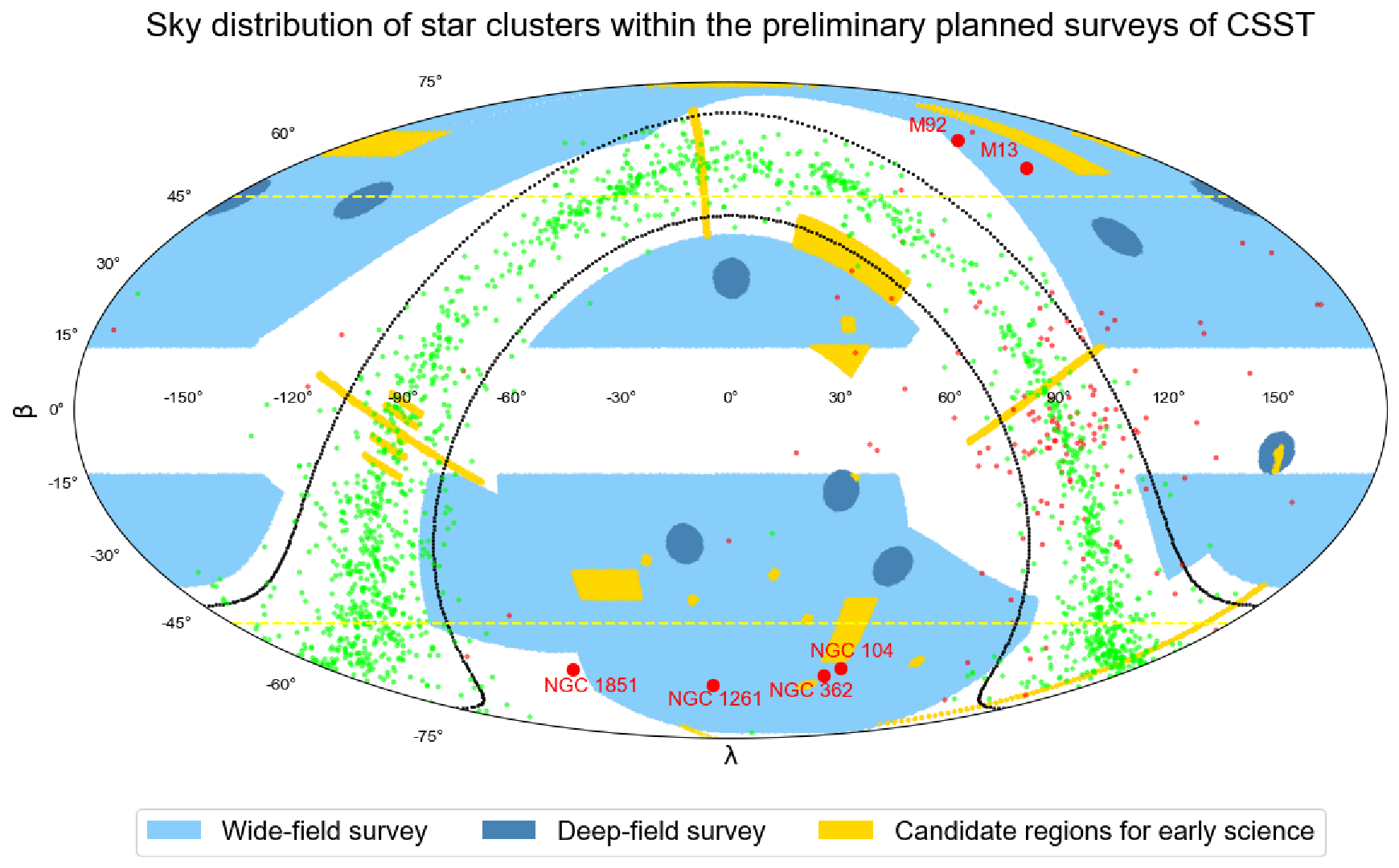}
        \caption{Distribution of the candidate star clusters overlaid on the preliminary survey regions of the CSST. The legend of this figure is consistent with that of Figure \ref{fig:all_sky}.}
        \label{fig:cluster_plannedsurveys}
    \end{figure}

    Figure \ref{fig:all_sky} illustrates the sky distribution of the six recommended globular clusters, overlaid on the \textit{Planck} dust extinction map. Figure \ref{fig:cluster_plannedsurveys} shows their positions relative to the preliminary planned surveys of the China Space Station Telescope (CSST), including the $\rm 17{,}500 ~ deg^2$ Wide-field Survey, the 400 $\rm deg^2$ Deep-field Survey, and other candidate regions designated for its early science \citep[see][]{Gong2025}. Among them, five clusters lie within the preliminary footprint of the Wide-field Survey, while NGC 1851 is located just outside these planned regions but is very close to the boundary of the Wide-field Survey area. We emphasize that these survey regions are still preliminary at the current stage and may be adjusted as the scientific priorities of the CSST early science program evolve.
    
	\subsection{Observational Visibility Analysis}
	
	Following the methodology outlined in Section \ref{s21}, we further calculate $D_{\rm obs}$, $H_{\rm obs}$, and $\sigma_{\rm frac}$ for the selected candidate star clusters. The detailed results are presented in Table \ref{table:info_6cl}. Our analysis reveals that the selected candidate clusters have an average annual observational visibility with the MSC of 236–253 observable days and 1254–1701 observable hours.
	
\begin{figure}[htbp]
		\centering
		\includegraphics[width=\textwidth]{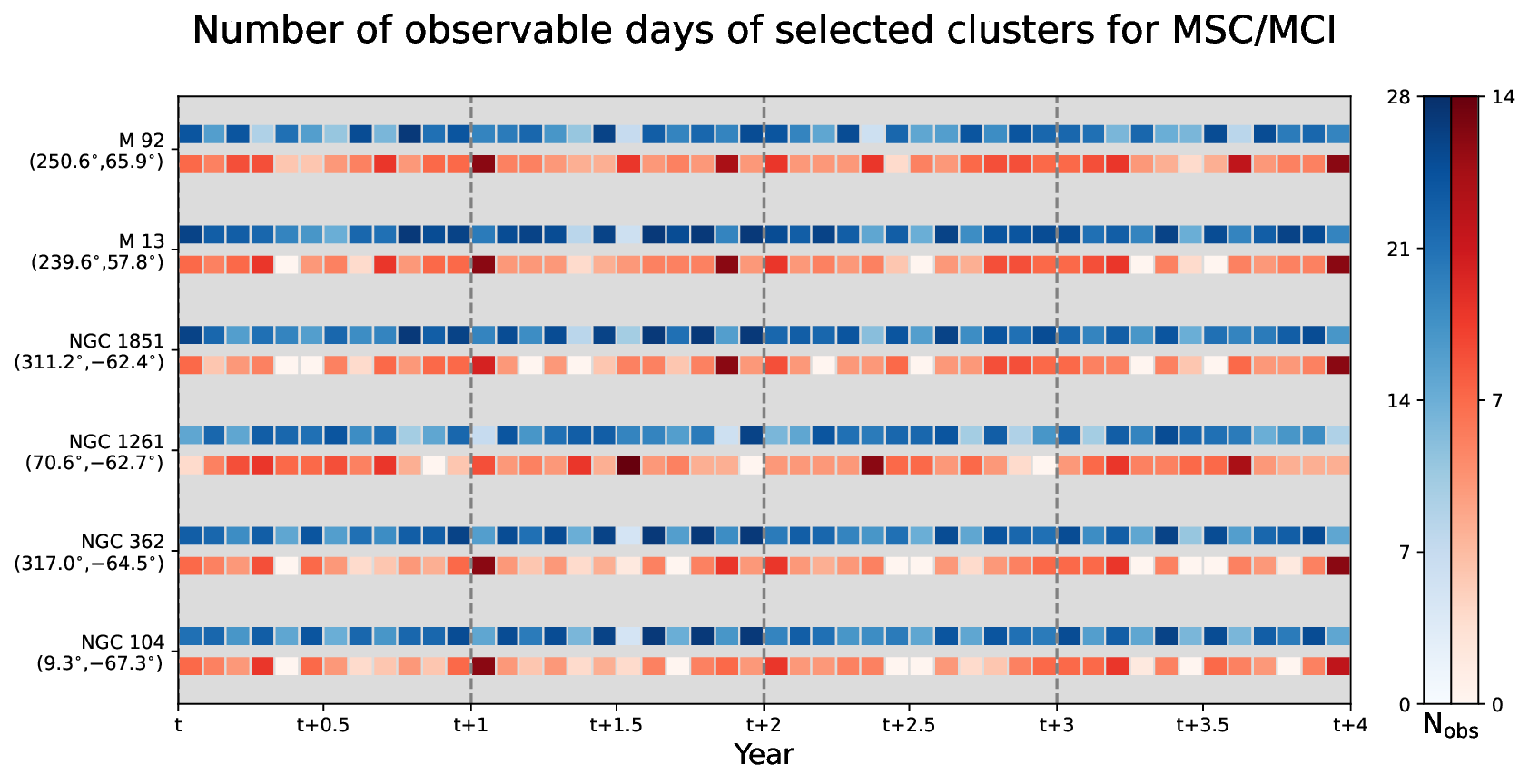}
		\caption{The number of observable days in each month for the six selected candidate star clusters in a four-year period. Each cell represents the number of observable days in a given month for one cluster and their brightness are related to the number of observational days (see the $\rm N_{obs}$ in the colorbar) in each month. Blue color for MSC and red color for MCI, respectively.}
		\label{fig:obsdays_6}
	\end{figure}

	Additionally, we analyze the monthly distribution of observable days over a four-year period. For the six selected star clusters, they have 6 to 28 observable days per month. This range provides sufficient flexibility for scheduling calibration observations throughout the year, with a minimum of six observable days per month to ensure regular calibration opportunities. However, for some clusters, the MCI experiences one or two months annually with very few or zero observable days (see Figure \ref{fig:obsdays_6}). This results from the MCI's visibility characteristics, whose observational visibility drops rapidly in specific months due to stronger lunar orbital influences on its assigned orbital segments. Therefore, this finding suggests that the on-orbit calibration strategy should be adjusted to avoid these periods of MCI invisibility.
	
	\begin{table}[htbp]
		\centering
		\begin{tabular}{ccccccccc}
		\toprule
		Name &  $\si{\lambda}$ ($\si{\degree}$) & $\si{\beta}$ ($\si{\degree}$) & $\overline{D_{\rm MSC}}$ & $\overline{D_{\rm MCI}}$ &$\overline{H_{\rm MSC}}$ & $\overline{H_{\rm MCI}}$& $\overline{\sigma_{\rm MSC}}$ & $\overline{\sigma_{\rm MCI}}$ \\ 
		\midrule
        M92 (NGC 6341)  & 234.3&  68.7& 243  & 63 & 1660.1 & 383.5 & 0.165 & 0.091  \\ 
        M13 (NGC 6205)  & 250.6&  65.9& 248  & 62 & 1700.5 & 390.2 & 0.160 & 0.095  \\ 
        NGC 104         & 311.2& -62.4& 248  & 73 & 1253.9 & 355.1 & 0.128 & 0.092  \\ 
        NGC 1851        &  70.6& -62.7& 253  & 60 & 1530.5 & 364.7 & 0.159 & 0.098  \\
        NGC 362         & 317.0& -64.5& 249  & 73 & 1269.8 & 354.4 & 0.131 & 0.090  \\ 
        NGC 1261        &   9.3& -67.3& 236  & 66 & 1344.8 & 342.0 & 0.165 & 0.086  \\ 
			\bottomrule
		\end{tabular}
		\caption{Observational visibility parameters of the six candidate star clusters. $\overline{D_{\rm MSC/MCI}}$, $\overline{H_{\rm MSC/MCI}}$, and $\overline{\sigma_{\rm MSC/MCI}}$ denote the average annual of $D_{\rm obs}$ (days), $H_{\rm obs}$ (hours), and $\sigma_{\rm frac}$ for the MSC and MCI, respectively.}	
		\label{table:info_6cl}
	\end{table}
	
    \subsection{Bright Star Contamination}

\begin{figure}[htbp]
        \centering
        \includegraphics[width=\textwidth]{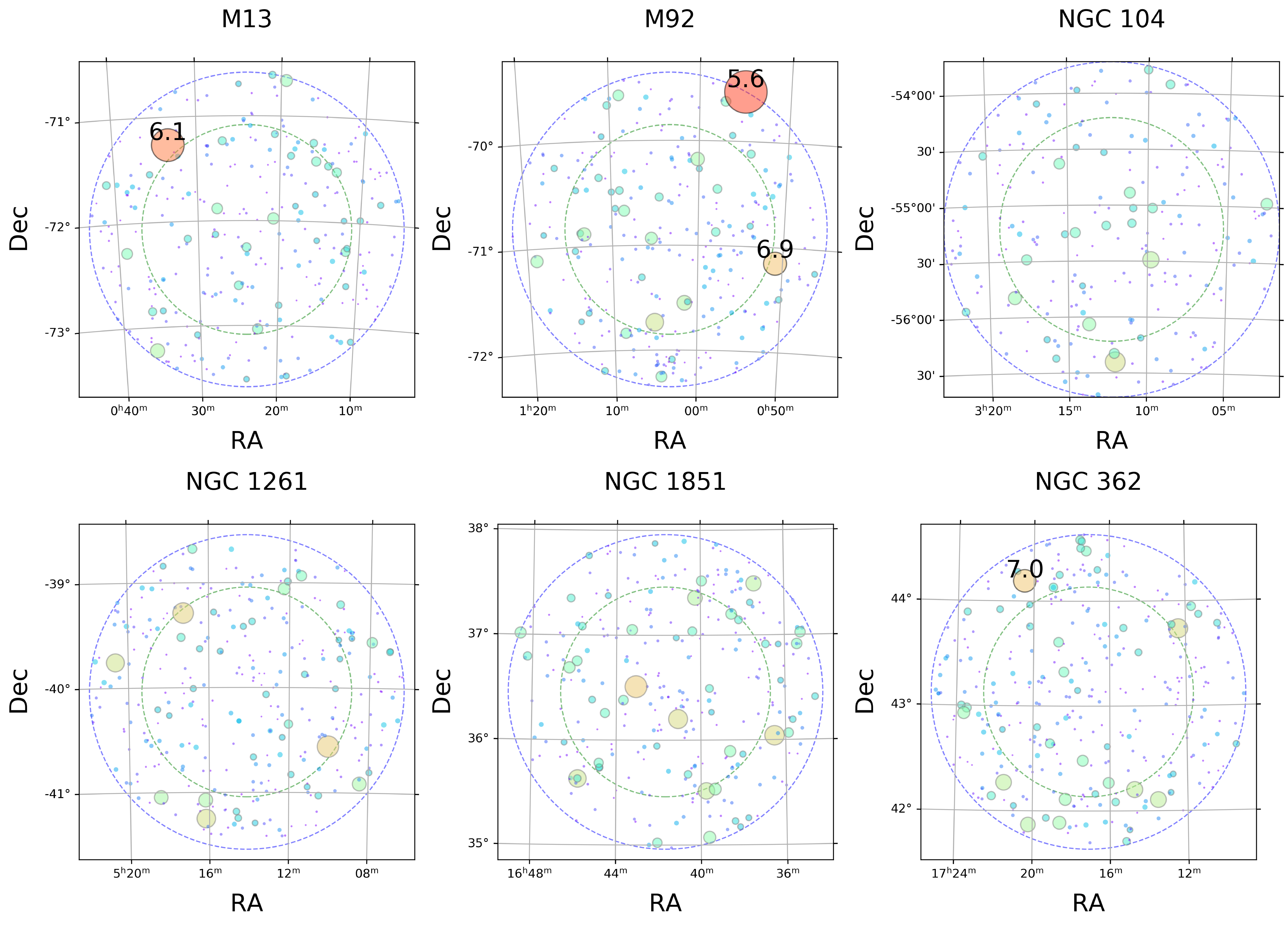}
        \caption{Sky distribution of Tycho-2 stars around the candidate star clusters. The cyan and blue dashed circles denote the regions with radii $r = \ang{1}$ and $r = \ang{1.5}$, respectively. The colors and sizes of the plotted stars are scaled according to their apparent magnitudes, with red representing bright stars and blue representing faint stars, respectively. Stars with $V \leq 10$ are overlaid with gray circular boundaries. Stars with $V \leq 7$ are overlaid with black circular boundaries and their corresponding $V$-band magnitudes are overlaid on top of each circle.}
        \label{fig:star_6}
    \end{figure}

    Figure \ref{fig:star_6} presents the sky distribution of Tycho-2 catalog stars in the fields surrounding the candidate star clusters. The fields of NGC 1261, NGC 1851 and NGC 6205 exhibit no bright stars of $V \le 7$ within a radius of $\ang{1.5}$. NGC 104, NGC 362 and NGC 6341 contain 1, 2, and 1 such stars ($V \le 7$) within $r=\ang{1.5}$, respectively. The observational strategies may require careful planning to avoid the contamination effects from these $V \le 7$ bright stars. Besides, none of these fields are contaminated by bright nebulae or extended galaxies based on visual inspection.
    
    \subsection{Extinction Mapping}
    
\begin{figure}[htbp]
		\centering
		\includegraphics[width=\textwidth]{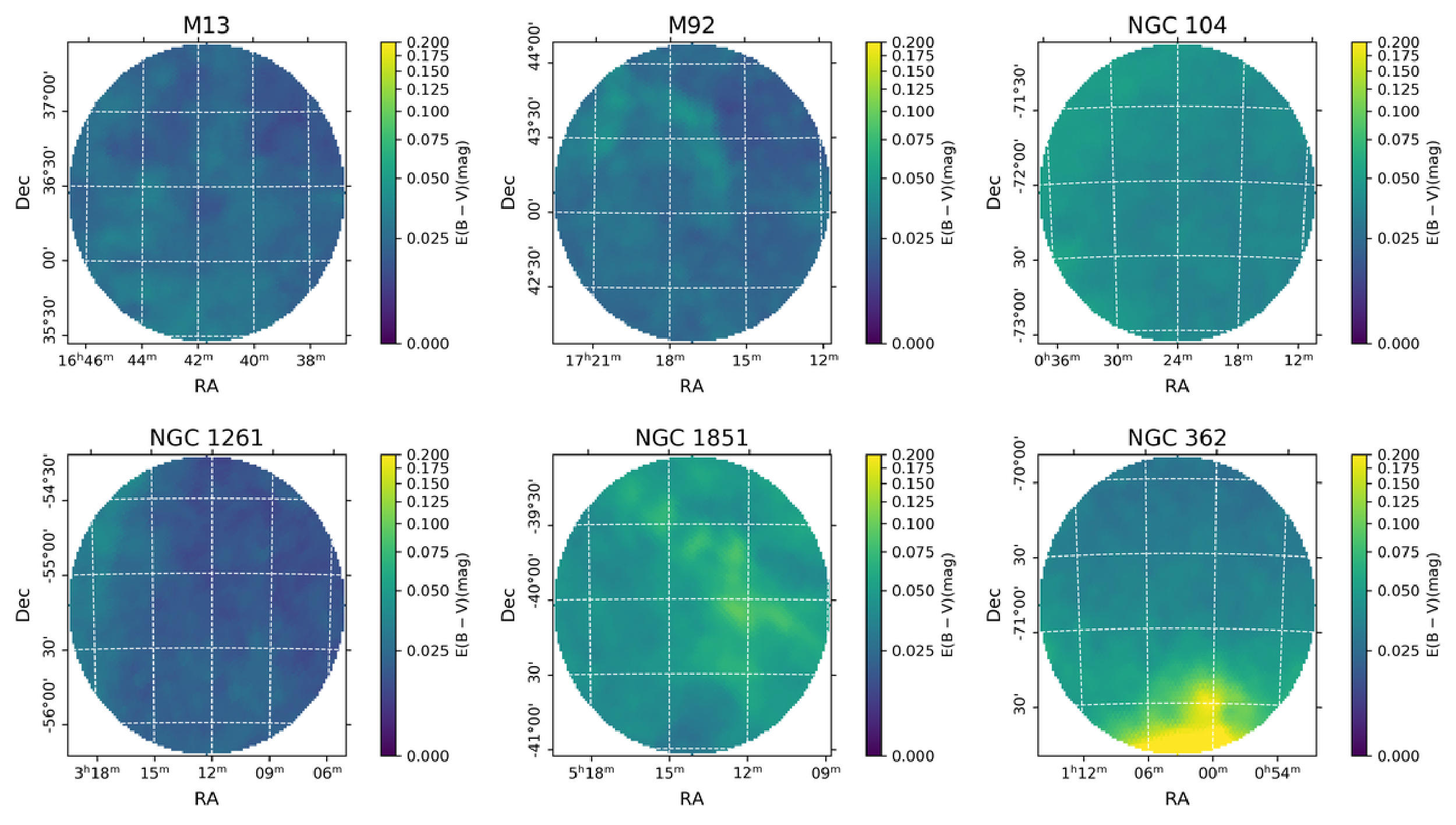}
		\caption{The \textit{Planck} dust extinction map around the selected candidate star clusters with a search radius of $r = \ang{1}$. The colors represent the $E(B - V)$ values from the \textit{Planck} map, as indicated by the color bar.}
		\label{fig:ebv_6}
	\end{figure}

    Figure \ref{fig:ebv_6} presents the dust extinction maps for the six candidate star clusters, generated using the same method described in Section \ref{s2.2}.  The $ E(B-V)_{\rm mean}$ values for these fields range from 0.025 to 0.054 (see Table \ref{table:cluster_details}), confirming that all six star clusters satisfy our selection criteria. However, 
    the map for NGC 362 reveals a region approximately \ang{1} to the south with a significantly higher extinction value of $E(B-V) > 0.15$. This localized anomaly elevates the $E(B-V)_{max}$ within the search radius to 0.441 mag (see Table \ref{table:cluster_details}) and is associated with the extended dust structure of the Small Magellanic Cloud (SMC). Despite the small spatial extent of this affected area, the observational strategy should still be designed to minimize its impact. For the remaining five candidate clusters, dust extinction does not significantly compromise the on-orbit calibration observations.

    \section{Conclusion}
    \label{s5}
        
        This work presents a systematic method for selecting optimal on-orbit calibration fields for the CSST imaging survey. We have developed and applied a two-stage screening framework and ultimately recommend six globular clusters.
        
        In the first stage of our selection strategy, we established two key spatial constraints by quantifying the key environmental factors critical for calibration field feasibility. Using \texttt{COSAT}, we performed an 11-year (2027--2038) all-sky visibility simulation for both the MSC and MCI, incorporating critical orbital constraints including Solar/Lunar/Earth avoidance and SAA exclusion. Our results reveal a critical threshold at an ecliptic latitude of $\lvert \beta \rvert = \ang{50}$: regions with $\lvert \beta \rvert \geq \ang{50}$ yield an  average annual of $\sim 240$ observable days for the MSC, with low monthly variability in visibility and cumulative annual observable hours exceeding $\SI{1300}{\hour}$. In contrast, regions below this threshold suffer a $\sim30\%$ decrease in observable days and experience significant seasonal gaps in visibility (2--3 consecutive months of non-visibility). Additionally, by combining the SFD and \textit{Planck} all-sky dust extinction maps and adopting a scientifically justified extinction threshold of $E(B-V) \leq 0.15$, we determined that galactic latitude constraint of $\lvert b \rvert \geq \ang{15}$ effectively mitigates interstellar extinction in CSST's observational bands. These two spatial constraints ($\lvert \beta \rvert \geq \ang{50}$, $\lvert b \rvert \geq \ang{15}$) form the fundamental feasibility criteria for CSST calibration fields, ensuring stable long-term observational access and minimal extinction contamination.
        
        Applying these spatial constraints to the 3006 star clusters in the Milky Way Star Clusters (MWSC) catalog yielded 29 candidate clusters (10 globular and 19 open clusters). In the second stage, we imposed three calibration-specific instrumental selection criteria and conducted a detailed quantitative evaluation of each candidate: (1) a stellar number density of $50 \leq \rho_{\rm star} \leq 1000$ stars arcmin$^{-2}$ (Gaia $G$-band, $16 \leq G \leq 22$), matching the CSST CCD chip size and the telescope's FOV; (2) strict avoidance of bright stars with $V \leq 7$ mag within a $\ang{1}$ radius to prevent detector saturation and stray light artifacts; (3) dual dust extinction constraints of $E(B-V)_{\rm cl} \leq 0.15$ (intrinsic cluster extinction) and $E(B-V)_{\rm mean} \leq 0.15$ (line-of-sight extinction derived from the \textit{Planck} dust map) within a $\ang{1}$ radius. Through this rigorous multi-parameter screening, we identified six globular clusters that satisfy all primary and secondary calibration requirements as the top-tier on-orbit calibration field candidates for CSST: M13 (NGC 6205), M92 (NGC 6341), NGC 104, NGC 362, NGC 1261, and NGC 1851.
        
        In-depth characterizations of the six globular clusters further validate their suitability for CSST calibration, and yield specific observational planning recommendations. Spatially, these clusters cover both the northern and southern celestial hemispheres, with ecliptic latitudes at $\ang{62.4} \leq \lvert \beta \rvert \leq \ang{68.7}$, ensuring balanced calibration coverage. For the MSC, their annual observable days range from 236 to 253, with monthly observable days varying between 6 and 28, providing sufficient flexibility for calibration scheduling. Stellar density analysis shows that all six clusters have chip-scale densities matching CSST calibration needs, with M13 and NGC 104 being the optimal candidates due to their high density and large angular size. While all candidates avoid bright stars ($V \leq 7$) within $\ang{1}$ radius, three clusters (NGC 104, NGC 362, M92) contain 1--2 bright stars within  $\ang{1.5}$ radius, an issue addressable by precise pointing. Dust extinction analysis reveals an average $E(B-V)$ of 0.025--0.054 for all six fields; the only potential extinction issue is a small high-extinction region ($E(B-V) > 0.15$) at a distance of $\ang{1}$ south of NGC 362, which can be avoided in specific observational designs. For the MCI, we note that some candidate star clusters exhibit one or two months each year with very few or even no observable days. This factor must be integrated into the CSST calibration scheduling strategy to prevent observational gaps. Notably, five of the clusters lie within the preliminary footprint of the CSST Wide-field Survey, with NGC 1851 adjacent to its boundary.
        
        This work presents the first comprehensive set of optimal calibration field candidates for the CSST imaging survey, accompanied by detailed characterizations of their observational and physical properties. Beyond its application to CSST, our two-stage selection methodology establishes a general framework for future space telescopes. By combining orbital dynamics, all-sky environmental analysis, and instrument-specific screening, it is particularly suited for wide-field imaging missions with UV/optical coverage.
        
       In the future, based on the analysis of this paper and combined with specific calibration requirements, we will optimize the design of calibration observation pointing. Accordingly, we will subsequently conduct simulations and data processing analysis, thus optimizing the on-orbit calibration scheme. This subsequent work will finalize the calibration field set, provide operational guidelines, and lay the groundwork for the scientific exploitation of the CSST survey data.
            	
	\begin{acknowledgements}
	This paper is funded by the National Key R\&D Program of China grants No. 2025YFF0511000. This work is also supported by the China Manned Space Project with grant No. CMS-CSST-2025-A19, No. CMS-CSST-2025-A08, and the CSST Scientific Data Processing and Analysis System of the China Manned Space Project.
	\end{acknowledgements}
	
	\bibliography{bibtex}

@ARTICLE{green2018,
       author = {{Green}, {Gregory M.}},
        title = "{dustmaps: A Python interface for maps of interstellar dust}",
      journal = {The Journal of Open Source Software},
         year = "2018",
        month = "Jun",
       volume = {3},
       number = {26},
        pages = {695},
          doi = {10.21105/joss.00695},
       adsurl = {https://ui.adsabs.harvard.edu/abs/2018JOSS....3..695G}
}

@ARTICLE{sfd,
       author = {{Schlegel}, David J. and {Finkbeiner}, Douglas P. and {Davis}, Marc},
        title = "{Maps of Dust Infrared Emission for Use in Estimation of Reddening and Cosmic Microwave Background Radiation Foregrounds}",
      journal = {\apj},
         year = 1998,
        month = jun,
       volume = {500},
       number = {2},
        pages = {525-553},
          doi = {10.1086/305772},
archivePrefix = {arXiv},
       eprint = {astro-ph/9710327},
 primaryClass = {astro-ph},
       adsurl = {https://ui.adsabs.harvard.edu/abs/1998ApJ...500..525S}
}

@ARTICLE{planck2016,
       author = {{Planck Collaboration} and {Aghanim}, N. and {Ashdown}, M. and {Aumont}, J. and {Baccigalupi}, C. and {Ballardini}, M. and {Banday}, A.~J. and {Barreiro}, R.~B. and {Bartolo}, N. and {Basak}, S. and {Benabed}, K. and {Bernard}, J. -P. and {Bersanelli}, M. and {Bielewicz}, P. and {Bonavera}, L. and {Bond}, J.~R. and {Borrill}, J. and {Bouchet}, F.~R. and {Boulanger}, F. and {Burigana}, C. and {Calabrese}, E. and {Cardoso}, J. -F. and {Carron}, J. and {Chiang}, H.~C. and {Colombo}, L.~P.~L. and {Comis}, B. and {Couchot}, F. and {Coulais}, A. and {Crill}, B.~P. and {Curto}, A. and {Cuttaia}, F. and {de Bernardis}, P. and {de Zotti}, G. and {Delabrouille}, J. and {Di Valentino}, E. and {Dickinson}, C. and {Diego}, J.~M. and {Dor{\'e}}, O. and {Douspis}, M. and {Ducout}, A. and {Dupac}, X. and {Dusini}, S. and {Elsner}, F. and {En{\ss}lin}, T.~A. and {Eriksen}, H.~K. and {Falgarone}, E. and {Fantaye}, Y. and {Finelli}, F. and {Forastieri}, F. and {Frailis}, M. and {Fraisse}, A.~A. and {Franceschi}, E. and {Frolov}, A. and {Galeotta}, S. and {Galli}, S. and {Ganga}, K. and {G{\'e}nova-Santos}, R.~T. and {Gerbino}, M. and {Ghosh}, T. and {Giraud-H{\'e}raud}, Y. and {Gonz{\'a}lez-Nuevo}, J. and {G{\'o}rski}, K.~M. and {Gruppuso}, A. and {Gudmundsson}, J.~E. and {Hansen}, F.~K. and {Helou}, G. and {Henrot-Versill{\'e}}, S. and {Herranz}, D. and {Hivon}, E. and {Huang}, Z. and {Jaffe}, A.~H. and {Jones}, W.~C. and {Keih{\"a}nen}, E. and {Keskitalo}, R. and {Kiiveri}, K. and {Kisner}, T.~S. and {Krachmalnicoff}, N. and {Kunz}, M. and {Kurki-Suonio}, H. and {Lamarre}, J. -M. and {Langer}, M. and {Lasenby}, A. and {Lattanzi}, M. and {Lawrence}, C.~R. and {Le Jeune}, M. and {Levrier}, F. and {Lilje}, P.~B. and {Lilley}, M. and {Lindholm}, V. and {L{\'o}pez-Caniego}, M. and {Ma}, Y. -Z. and {Mac{\'\i}as-P{\'e}rez}, J.~F. and {Maggio}, G. and {Maino}, D. and {Mandolesi}, N. and {Mangilli}, A. and {Maris}, M. and {Martin}, P.~G. and {Mart{\'\i}nez-Gonz{\'a}lez}, E. and {Matarrese}, S. and {Mauri}, N. and {McEwen}, J.~D. and {Melchiorri}, A. and {Mennella}, A. and {Migliaccio}, M. and {Miville-Desch{\^e}nes}, M. -A. and {Molinari}, D. and {Moneti}, A. and {Montier}, L. and {Morgante}, G. and {Moss}, A. and {Natoli}, P. and {Oxborrow}, C.~A. and {Pagano}, L. and {Paoletti}, D. and {Patanchon}, G. and {Perdereau}, O. and {Perotto}, L. and {Pettorino}, V. and {Piacentini}, F. and {Plaszczynski}, S. and {Polastri}, L. and {Polenta}, G. and {Puget}, J. -L. and {Rachen}, J.~P. and {Racine}, B. and {Reinecke}, M. and {Remazeilles}, M. and {Renzi}, A. and {Rocha}, G. and {Rosset}, C. and {Rossetti}, M. and {Roudier}, G. and {Rubi{\~n}o-Mart{\'\i}n}, J.~A. and {Ruiz-Granados}, B. and {Salvati}, L. and {Sandri}, M. and {Savelainen}, M. and {Scott}, D. and {Sirignano}, C. and {Sirri}, G. and {Soler}, J.~D. and {Spencer}, L.~D. and {Suur-Uski}, A. -S. and {Tauber}, J.~A. and {Tavagnacco}, D. and {Tenti}, M. and {Toffolatti}, L. and {Tomasi}, M. and {Tristram}, M. and {Trombetti}, T. and {Valiviita}, J. and {Van Tent}, F. and {Vielva}, P. and {Villa}, F. and {Vittorio}, N. and {Wandelt}, B.~D. and {Wehus}, I.~K. and {Zacchei}, A. and {Zonca}, A.},
        title = "{Planck intermediate results. XLVIII. Disentangling Galactic dust emission and cosmic infrared background anisotropies}",
      journal = {\aap},
         year = 2016,
        month = dec,
       volume = {596},
          eid = {A109},
        pages = {A109},
          doi = {10.1051/0004-6361/201629022},
archivePrefix = {arXiv},
       eprint = {1605.09387},
 primaryClass = {astro-ph.CO},
       adsurl = {https://ui.adsabs.harvard.edu/abs/2016A&A...596A.109P}
}

@ARTICLE{hog2000,
       author = {{H{\o}g}, E. and {Fabricius}, C. and {Makarov}, V.~V. and {Urban}, S. and {Corbin}, T. and {Wycoff}, G. and {Bastian}, U. and {Schwekendiek}, P. and {Wicenec}, A.},
        title = "{The Tycho-2 catalogue of the 2.5 million brightest stars}",
      journal = {\aap},
         year = 2000,
        month = mar,
       volume = {355},
        pages = {L27-L30},
       adsurl = {https://ui.adsabs.harvard.edu/abs/2000A&A...355L..27H}
}

@ARTICLE{astroquery,
   author = {{Ginsburg}, A. and {Sip{\H o}cz}, B.~M. and {Brasseur}, C.~E. and
	{Cowperthwaite}, P.~S. and {Craig}, M.~W. and {Deil}, C. and
	{Guillochon}, J. and {Guzman}, G. and {Liedtke}, S. and {Lian Lim}, P. and
	{Lockhart}, K.~E. and {Mommert}, M. and {Morris}, B.~M. and
	{Norman}, H. and {Parikh}, M. and {Persson}, M.~V. and {Robitaille}, T.~P. and
	{Segovia}, J.-C. and {Singer}, L.~P. and {Tollerud}, E.~J. and
	{de Val-Borro}, M. and {Valtchanov}, I. and {Woillez}, J. and
	{The Astroquery collaboration} and {a subset of the astropy collaboration}
	},
    title = "{astroquery: An Astronomical Web-querying Package in Python}",
  journal = {\aj},
archivePrefix = "arXiv",
   eprint = {1901.04520},
 primaryClass = "astro-ph.IM",
     year = 2019,
    month = mar,
   volume = 157,
      eid = {98},
    pages = {98},
      doi = {10.3847/1538-3881/aafc33},
   adsurl = {https://adsabs.harvard.edu/abs/2019AJ....157...98G}
}

@ARTICLE{gaia2016,
       author = {{Gaia Collaboration} and {Prusti}, T. and {de Bruijne}, J.~H.~J. and {Brown}, A.~G.~A. and {Vallenari}, A. and {Babusiaux}, C. and {Bailer-Jones}, C.~A.~L. and {Bastian}, U. and {Biermann}, M. and {Evans}, D.~W. and {Eyer}, L. and {Jansen}, F. and {Jordi}, C. and {Klioner}, S.~A. and {Lammers}, U. and {Lindegren}, L. and {Luri}, X. and {Mignard}, F. and {Milligan}, D.~J. and {Panem}, C. and {Poinsignon}, V. and {Pourbaix}, D. and {Randich}, S. and {Sarri}, G. and {Sartoretti}, P. and {Siddiqui}, H.~I. and {Soubiran}, C. and {Valette}, V. and {van Leeuwen}, F. and {Walton}, N.~A. and {Aerts}, C. and {Arenou}, F. and {Cropper}, M. and {Drimmel}, R. and {H{\o}g}, E. and {Katz}, D. and {Lattanzi}, M.~G. and {O'Mullane}, W. and {Grebel}, E.~K. and {Holland}, A.~D. and {Huc}, C. and {Passot}, X. and {Bramante}, L. and {Cacciari}, C. and {Casta{\~n}eda}, J. and {Chaoul}, L. and {Cheek}, N. and {De Angeli}, F. and {Fabricius}, C. and {Guerra}, R. and {Hern{\'a}ndez}, J. and {Jean-Antoine-Piccolo}, A. and {Masana}, E. and {Messineo}, R. and {Mowlavi}, N. and {Nienartowicz}, K. and {Ord{\'o}{\~n}ez-Blanco}, D. and {Panuzzo}, P. and {Portell}, J. and {Richards}, P.~J. and {Riello}, M. and {Seabroke}, G.~M. and {Tanga}, P. and {Th{\'e}venin}, F. and {Torra}, J. and {Els}, S.~G. and {Gracia-Abril}, G. and {Comoretto}, G. and {Garcia-Reinaldos}, M. and {Lock}, T. and {Mercier}, E. and {Altmann}, M. and {Andrae}, R. and {Astraatmadja}, T.~L. and {Bellas-Velidis}, I. and {Benson}, K. and {Berthier}, J. and {Blomme}, R. and {Busso}, G. and {Carry}, B. and {Cellino}, A. and {Clementini}, G. and {Cowell}, S. and {Creevey}, O. and {Cuypers}, J. and {Davidson}, M. and {De Ridder}, J. and {de Torres}, A. and {Delchambre}, L. and {Dell'Oro}, A. and {Ducourant}, C. and {Fr{\'e}mat}, Y. and {Garc{\'\i}a-Torres}, M. and {Gosset}, E. and {Halbwachs}, J. -L. and {Hambly}, N.~C. and {Harrison}, D.~L. and {Hauser}, M. and {Hestroffer}, D. and {Hodgkin}, S.~T. and {Huckle}, H.~E. and {Hutton}, A. and {Jasniewicz}, G. and {Jordan}, S. and {Kontizas}, M. and {Korn}, A.~J. and {Lanzafame}, A.~C. and {Manteiga}, M. and {Moitinho}, A. and {Muinonen}, K. and {Osinde}, J. and {Pancino}, E. and {Pauwels}, T. and {Petit}, J. -M. and {Recio-Blanco}, A. and {Robin}, A.~C. and {Sarro}, L.~M. and {Siopis}, C. and {Smith}, M. and {Smith}, K.~W. and {Sozzetti}, A. and {Thuillot}, W. and {van Reeven}, W. and {Viala}, Y. and {Abbas}, U. and {Abreu Aramburu}, A. and {Accart}, S. and {Aguado}, J.~J. and {Allan}, P.~M. and {Allasia}, W. and {Altavilla}, G. and {{\'A}lvarez}, M.~A. and {Alves}, J. and {Anderson}, R.~I. and {Andrei}, A.~H. and {Anglada Varela}, E. and {Antiche}, E. and {Antoja}, T. and {Ant{\'o}n}, S. and {Arcay}, B. and {Atzei}, A. and {Ayache}, L. and {Bach}, N. and {Baker}, S.~G. and {Balaguer-N{\'u}{\~n}ez}, L. and {Barache}, C. and {Barata}, C. and {Barbier}, A. and {Barblan}, F. and {Baroni}, M. and {Barrado y Navascu{\'e}s}, D. and {Barros}, M. and {Barstow}, M.~A. and {Becciani}, U. and {Bellazzini}, M. and {Bellei}, G. and {Bello Garc{\'\i}a}, A. and {Belokurov}, V. and {Bendjoya}, P. and {Berihuete}, A. and {Bianchi}, L. and {Bienaym{\'e}}, O. and {Billebaud}, F. and {Blagorodnova}, N. and {Blanco-Cuaresma}, S. and {Boch}, T. and {Bombrun}, A. and {Borrachero}, R. and {Bouquillon}, S. and {Bourda}, G. and {Bouy}, H. and {Bragaglia}, A. and {Breddels}, M.~A. and {Brouillet}, N. and {Br{\"u}semeister}, T. and {Bucciarelli}, B. and {Budnik}, F. and {Burgess}, P. and {Burgon}, R. and {Burlacu}, A. and {Busonero}, D. and {Buzzi}, R. and {Caffau}, E. and {Cambras}, J. and {Campbell}, H. and {Cancelliere}, R. and {Cantat-Gaudin}, T. and {Carlucci}, T. and {Carrasco}, J.~M. and {Castellani}, M. and {Charlot}, P. and {Charnas}, J. and {Charvet}, P. and {Chassat}, F. and {Chiavassa}, A. and {Clotet}, M. and {Cocozza}, G. and {Collins}, R.~S. and {Collins}, P. and {Costigan}, G.},
        title = "{The Gaia mission}",
      journal = {\aap},
         year = 2016,
        month = nov,
       volume = {595},
          eid = {A1},
        pages = {A1},
          doi = {10.1051/0004-6361/201629272},
archivePrefix = {arXiv},
       eprint = {1609.04153},
 primaryClass = {astro-ph.IM},
       adsurl = {https://ui.adsabs.harvard.edu/abs/2016A&A...595A...1G}
}

@ARTICLE{gaia2023,
       author = {{Gaia Collaboration} and {Vallenari}, A. and {Brown}, A.~G.~A. and {Prusti}, T. and {de Bruijne}, J.~H.~J. and {Arenou}, F. and {Babusiaux}, C. and {Biermann}, M. and {Creevey}, O.~L. and {Ducourant}, C. and {Evans}, D.~W. and {Eyer}, L. and {Guerra}, R. and {Hutton}, A. and {Jordi}, C. and {Klioner}, S.~A. and {Lammers}, U.~L. and {Lindegren}, L. and {Luri}, X. and {Mignard}, F. and {Panem}, C. and {Pourbaix}, D. and {Randich}, S. and {Sartoretti}, P. and {Soubiran}, C. and {Tanga}, P. and {Walton}, N.~A. and {Bailer-Jones}, C.~A.~L. and {Bastian}, U. and {Drimmel}, R. and {Jansen}, F. and {Katz}, D. and {Lattanzi}, M.~G. and {van Leeuwen}, F. and {Bakker}, J. and {Cacciari}, C. and {Casta{\~n}eda}, J. and {De Angeli}, F. and {Fabricius}, C. and {Fouesneau}, M. and {Fr{\'e}mat}, Y. and {Galluccio}, L. and {Guerrier}, A. and {Heiter}, U. and {Masana}, E. and {Messineo}, R. and {Mowlavi}, N. and {Nicolas}, C. and {Nienartowicz}, K. and {Pailler}, F. and {Panuzzo}, P. and {Riclet}, F. and {Roux}, W. and {Seabroke}, G.~M. and {Sordo}, R. and {Th{\'e}venin}, F. and {Gracia-Abril}, G. and {Portell}, J. and {Teyssier}, D. and {Altmann}, M. and {Andrae}, R. and {Audard}, M. and {Bellas-Velidis}, I. and {Benson}, K. and {Berthier}, J. and {Blomme}, R. and {Burgess}, P.~W. and {Busonero}, D. and {Busso}, G. and {C{\'a}novas}, H. and {Carry}, B. and {Cellino}, A. and {Cheek}, N. and {Clementini}, G. and {Damerdji}, Y. and {Davidson}, M. and {de Teodoro}, P. and {Nu{\~n}ez Campos}, M. and {Delchambre}, L. and {Dell'Oro}, A. and {Esquej}, P. and {Fern{\'a}ndez-Hern{\'a}ndez}, J. and {Fraile}, E. and {Garabato}, D. and {Garc{\'\i}a-Lario}, P. and {Gosset}, E. and {Haigron}, R. and {Halbwachs}, J. -L. and {Hambly}, N.~C. and {Harrison}, D.~L. and {Hern{\'a}ndez}, J. and {Hestroffer}, D. and {Hodgkin}, S.~T. and {Holl}, B. and {Jan{\ss}en}, K. and {Jevardat de Fombelle}, G. and {Jordan}, S. and {Krone-Martins}, A. and {Lanzafame}, A.~C. and {L{\"o}ffler}, W. and {Marchal}, O. and {Marrese}, P.~M. and {Moitinho}, A. and {Muinonen}, K. and {Osborne}, P. and {Pancino}, E. and {Pauwels}, T. and {Recio-Blanco}, A. and {Reyl{\'e}}, C. and {Riello}, M. and {Rimoldini}, L. and {Roegiers}, T. and {Rybizki}, J. and {Sarro}, L.~M. and {Siopis}, C. and {Smith}, M. and {Sozzetti}, A. and {Utrilla}, E. and {van Leeuwen}, M. and {Abbas}, U. and {{\'A}brah{\'a}m}, P. and {Abreu Aramburu}, A. and {Aerts}, C. and {Aguado}, J.~J. and {Ajaj}, M. and {Aldea-Montero}, F. and {Altavilla}, G. and {{\'A}lvarez}, M.~A. and {Alves}, J. and {Anders}, F. and {Anderson}, R.~I. and {Anglada Varela}, E. and {Antoja}, T. and {Baines}, D. and {Baker}, S.~G. and {Balaguer-N{\'u}{\~n}ez}, L. and {Balbinot}, E. and {Balog}, Z. and {Barache}, C. and {Barbato}, D. and {Barros}, M. and {Barstow}, M.~A. and {Bartolom{\'e}}, S. and {Bassilana}, J. -L. and {Bauchet}, N. and {Becciani}, U. and {Bellazzini}, M. and {Berihuete}, A. and {Bernet}, M. and {Bertone}, S. and {Bianchi}, L. and {Binnenfeld}, A. and {Blanco-Cuaresma}, S. and {Blazere}, A. and {Boch}, T. and {Bombrun}, A. and {Bossini}, D. and {Bouquillon}, S. and {Bragaglia}, A. and {Bramante}, L. and {Breedt}, E. and {Bressan}, A. and {Brouillet}, N. and {Brugaletta}, E. and {Bucciarelli}, B. and {Burlacu}, A. and {Butkevich}, A.~G. and {Buzzi}, R. and {Caffau}, E. and {Cancelliere}, R. and {Cantat-Gaudin}, T. and {Carballo}, R. and {Carlucci}, T. and {Carnerero}, M.~I. and {Carrasco}, J.~M. and {Casamiquela}, L. and {Castellani}, M. and {Castro-Ginard}, A. and {Chaoul}, L. and {Charlot}, P. and {Chemin}, L. and {Chiaramida}, V. and {Chiavassa}, A. and {Chornay}, N. and {Comoretto}, G. and {Contursi}, G. and {Cooper}, W.~J. and {Cornez}, T. and {Cowell}, S. and {Crifo}, F. and {Cropper}, M. and {Crosta}, M. and {Crowley}, C. and {Dafonte}, C. and {Dapergolas}, A. and {David}, M. and {David}, P. and {de Laverny}, P. and {De Luise}, F. and {De March}, R.},
        title = "{Gaia Data Release 3. Summary of the content and survey properties}",
      journal = {\aap},
         year = 2023,
        month = jun,
       volume = {674},
          eid = {A1},
        pages = {A1},
          doi = {10.1051/0004-6361/202243940},
archivePrefix = {arXiv},
       eprint = {2208.00211},
 primaryClass = {astro-ph.GA},
       adsurl = {https://ui.adsabs.harvard.edu/abs/2023A&A...674A...1G}
}

@ARTICLE{kharchenko2012,
       author = {{Kharchenko}, N.~V. and {Piskunov}, A.~E. and {Schilbach}, E. and {R{\"o}ser}, S. and {Scholz}, R. -D.},
        title = "{Global survey of star clusters in the Milky Way. I. The pipeline and fundamental parameters in the second quadrant}",
      journal = {\aap},
         year = 2012,
        month = jul,
       volume = {543},
          eid = {A156},
        pages = {A156},
          doi = {10.1051/0004-6361/201118708},
archivePrefix = {arXiv},
       eprint = {1207.4001},
 primaryClass = {astro-ph.GA},
       adsurl = {https://ui.adsabs.harvard.edu/abs/2012A&A...543A.156K}
}

@ARTICLE{zhan2021,
author = {Zhan, H},
year = 2021,
title = {The wide-field multiband imaging and slitless spectroscopy survey to be carried out by the Survey Space Telescope of China Manned Space Program},
journal={ Chinese Science Bulletin},
volume =  {66},
pages =  {1290},
doi={https://doi.org/10.1360/TB-2021-0016}
}

@article{astropy:2013,
	Adsurl = {http://adsabs.harvard.edu/abs/2013A%26A...558A..33A},
	Archiveprefix = {arXiv},
	Author = {{Astropy Collaboration} and {Robitaille}, T.~P. and {Tollerud}, E.~J. and {Greenfield}, P. and {Droettboom}, M. and {Bray}, E. and {Aldcroft}, T. and {Davis}, M. and {Ginsburg}, A. and {Price-Whelan}, A.~M. and {Kerzendorf}, W.~E. and {Conley}, A. and {Crighton}, N. and {Barbary}, K. and {Muna}, D. and {Ferguson}, H. and {Grollier}, F. and {Parikh}, M.~M. and {Nair}, P.~H. and {Unther}, H.~M. and {Deil}, C. and {Woillez}, J. and {Conseil}, S. and {Kramer}, R. and {Turner}, J.~E.~H. and {Singer}, L. and {Fox}, R. and {Weaver}, B.~A. and {Zabalza}, V. and {Edwards}, Z.~I. and {Azalee Bostroem}, K. and {Burke}, D.~J. and {Casey}, A.~R. and {Crawford}, S.~M. and {Dencheva}, N. and {Ely}, J. and {Jenness}, T. and {Labrie}, K. and {Lim}, P.~L. and {Pierfederici}, F. and {Pontzen}, A. and {Ptak}, A. and {Refsdal}, B. and {Servillat}, M. and {Streicher}, O.},
	Doi = {10.1051/0004-6361/201322068},
	Eid = {A33},
	Eprint = {1307.6212},
	Journal = {\aap},
	Month = oct,
	Pages = {A33},
	Primaryclass = {astro-ph.IM},
	Title = {{Astropy: A community Python package for astronomy}},
	Volume = 558,
	Year = 2013}

@ARTICLE{astropy:2018,
	author = {{Astropy Collaboration} and {Price-Whelan}, A.~M. and
	{Sip{\H{o}}cz}, B.~M. and {G{\"u}nther}, H.~M. and {Lim}, P.~L. and
	{Crawford}, S.~M. and {Conseil}, S. and {Shupe}, D.~L. and
	{Craig}, M.~W. and {Dencheva}, N. and {Ginsburg}, A. and {Vand
	erPlas}, J.~T. and {Bradley}, L.~D. and {P{\'e}rez-Su{\'a}rez}, D. and
	{de Val-Borro}, M. and {Aldcroft}, T.~L. and {Cruz}, K.~L. and
	{Robitaille}, T.~P. and {Tollerud}, E.~J. and {Ardelean}, C. and
	{Babej}, T. and {Bach}, Y.~P. and {Bachetti}, M. and {Bakanov}, A.~V. and
	{Bamford}, S.~P. and {Barentsen}, G. and {Barmby}, P. and
	{Baumbach}, A. and {Berry}, K.~L. and {Biscani}, F. and {Boquien}, M. and
	{Bostroem}, K.~A. and {Bouma}, L.~G. and {Brammer}, G.~B. and
	{Bray}, E.~M. and {Breytenbach}, H. and {Buddelmeijer}, H. and
	{Burke}, D.~J. and {Calderone}, G. and {Cano Rodr{\'\i}guez}, J.~L. and
	{Cara}, M. and {Cardoso}, J.~V.~M. and {Cheedella}, S. and {Copin}, Y. and
	{Corrales}, L. and {Crichton}, D. and {D'Avella}, D. and {Deil}, C. and
	{Depagne}, {\'E}. and {Dietrich}, J.~P. and {Donath}, A. and
	{Droettboom}, M. and {Earl}, N. and {Erben}, T. and {Fabbro}, S. and
	{Ferreira}, L.~A. and {Finethy}, T. and {Fox}, R.~T. and
	{Garrison}, L.~H. and {Gibbons}, S.~L.~J. and {Goldstein}, D.~A. and
	{Gommers}, R. and {Greco}, J.~P. and {Greenfield}, P. and
	{Groener}, A.~M. and {Grollier}, F. and {Hagen}, A. and {Hirst}, P. and
	{Homeier}, D. and {Horton}, A.~J. and {Hosseinzadeh}, G. and {Hu}, L. and
	{Hunkeler}, J.~S. and {Ivezi{\'c}}, {\v{Z}}. and {Jain}, A. and
	{Jenness}, T. and {Kanarek}, G. and {Kendrew}, S. and {Kern}, N.~S. and
	{Kerzendorf}, W.~E. and {Khvalko}, A. and {King}, J. and {Kirkby}, D. and
	{Kulkarni}, A.~M. and {Kumar}, A. and {Lee}, A. and {Lenz}, D. and
	{Littlefair}, S.~P. and {Ma}, Z. and {Macleod}, D.~M. and
	{Mastropietro}, M. and {McCully}, C. and {Montagnac}, S. and
	{Morris}, B.~M. and {Mueller}, M. and {Mumford}, S.~J. and {Muna}, D. and
	{Murphy}, N.~A. and {Nelson}, S. and {Nguyen}, G.~H. and
	{Ninan}, J.~P. and {N{\"o}the}, M. and {Ogaz}, S. and {Oh}, S. and
	{Parejko}, J.~K. and {Parley}, N. and {Pascual}, S. and {Patil}, R. and
	{Patil}, A.~A. and {Plunkett}, A.~L. and {Prochaska}, J.~X. and
	{Rastogi}, T. and {Reddy Janga}, V. and {Sabater}, J. and
	{Sakurikar}, P. and {Seifert}, M. and {Sherbert}, L.~E. and
	{Sherwood-Taylor}, H. and {Shih}, A.~Y. and {Sick}, J. and
	{Silbiger}, M.~T. and {Singanamalla}, S. and {Singer}, L.~P. and
	{Sladen}, P.~H. and {Sooley}, K.~A. and {Sornarajah}, S. and
	{Streicher}, O. and {Teuben}, P. and {Thomas}, S.~W. and
	{Tremblay}, G.~R. and {Turner}, J.~E.~H. and {Terr{\'o}n}, V. and
	{van Kerkwijk}, M.~H. and {de la Vega}, A. and {Watkins}, L.~L. and
	{Weaver}, B.~A. and {Whitmore}, J.~B. and {Woillez}, J. and
	{Zabalza}, V. and {Astropy Contributors}},
	title = "{The Astropy Project: Building an Open-science Project and Status of the v2.0 Core Package}",
	journal = {\aj},
	year = 2018,
	month = sep,
	volume = {156},
	number = {3},
	eid = {123},
	pages = {123},
	doi = {10.3847/1538-3881/aabc4f},
	archivePrefix = {arXiv},
	eprint = {1801.02634},
	primaryClass = {astro-ph.IM},
	adsurl = {https://ui.adsabs.harvard.edu/abs/2018AJ....156..123A}
}

@ARTICLE{astropy:2022,
	author = {{Astropy Collaboration} and {Price-Whelan}, Adrian M. and {Lim}, Pey Lian and {Earl}, Nicholas and {Starkman}, Nathaniel and {Bradley}, Larry and {Shupe}, David L. and {Patil}, Aarya A. and {Corrales}, Lia and {Brasseur}, C.~E. and {N{"o}the}, Maximilian and {Donath}, Axel and {Tollerud}, Erik and {Morris}, Brett M. and {Ginsburg}, Adam and {Vaher}, Eero and {Weaver}, Benjamin A. and {Tocknell}, James and {Jamieson}, William and {van Kerkwijk}, Marten H. and {Robitaille}, Thomas P. and {Merry}, Bruce and {Bachetti}, Matteo and {G{"u}nther}, H. Moritz and {Aldcroft}, Thomas L. and {Alvarado-Montes}, Jaime A. and {Archibald}, Anne M. and {B{'o}di}, Attila and {Bapat}, Shreyas and {Barentsen}, Geert and {Baz{'a}n}, Juanjo and {Biswas}, Manish and {Boquien}, M{'e}d{'e}ric and {Burke}, D.~J. and {Cara}, Daria and {Cara}, Mihai and {Conroy}, Kyle E. and {Conseil}, Simon and {Craig}, Matthew W. and {Cross}, Robert M. and {Cruz}, Kelle L. and {D'Eugenio}, Francesco and {Dencheva}, Nadia and {Devillepoix}, Hadrien A.~R. and {Dietrich}, J{"o}rg P. and {Eigenbrot}, Arthur Davis and {Erben}, Thomas and {Ferreira}, Leonardo and {Foreman-Mackey}, Daniel and {Fox}, Ryan and {Freij}, Nabil and {Garg}, Suyog and {Geda}, Robel and {Glattly}, Lauren and {Gondhalekar}, Yash and {Gordon}, Karl D. and {Grant}, David and {Greenfield}, Perry and {Groener}, Austen M. and {Guest}, Steve and {Gurovich}, Sebastian and {Handberg}, Rasmus and {Hart}, Akeem and {Hatfield-Dodds}, Zac and {Homeier}, Derek and {Hosseinzadeh}, Griffin and {Jenness}, Tim and {Jones}, Craig K. and {Joseph}, Prajwel and {Kalmbach}, J. Bryce and {Karamehmetoglu}, Emir and {Ka{l}uszy{'n}ski}, Miko{l}aj and {Kelley}, Michael S.~P. and {Kern}, Nicholas and {Kerzendorf}, Wolfgang E. and {Koch}, Eric W. and {Kulumani}, Shankar and {Lee}, Antony and {Ly}, Chun and {Ma}, Zhiyuan and {MacBride}, Conor and {Maljaars}, Jakob M. and {Muna}, Demitri and {Murphy}, N.~A. and {Norman}, Henrik and {O'Steen}, Richard and {Oman}, Kyle A. and {Pacifici}, Camilla and {Pascual}, Sergio and {Pascual-Granado}, J. and {Patil}, Rohit R. and {Perren}, Gabriel I. and {Pickering}, Timothy E. and {Rastogi}, Tanuj and {Roulston}, Benjamin R. and {Ryan}, Daniel F. and {Rykoff}, Eli S. and {Sabater}, Jose and {Sakurikar}, Parikshit and {Salgado}, Jes{'u}s and {Sanghi}, Aniket and {Saunders}, Nicholas and {Savchenko}, Volodymyr and {Schwardt}, Ludwig and {Seifert-Eckert}, Michael and {Shih}, Albert Y. and {Jain}, Anany Shrey and {Shukla}, Gyanendra and {Sick}, Jonathan and {Simpson}, Chris and {Singanamalla}, Sudheesh and {Singer}, Leo P. and {Singhal}, Jaladh and {Sinha}, Manodeep and {Sip{H{o}}cz}, Brigitta M. and {Spitler}, Lee R. and {Stansby}, David and {Streicher}, Ole and {{{S}}umak}, Jani and {Swinbank}, John D. and {Taranu}, Dan S. and {Tewary}, Nikita and {Tremblay}, Grant R. and {Val-Borro}, Miguel de and {Van Kooten}, Samuel J. and {Vasovi{'c}}, Zlatan and {Verma}, Shresth and {de Miranda Cardoso}, Jos{'e} Vin{'i}cius and {Williams}, Peter K.~G. and {Wilson}, Tom J. and {Winkel}, Benjamin and {Wood-Vasey}, W.~M. and {Xue}, Rui and {Yoachim}, Peter and {Zhang}, Chen and {Zonca}, Andrea and {Astropy Project Contributors}},
	title = "{The Astropy Project: Sustaining and Growing a Community-oriented Open-source Project and the Latest Major Release (v5.0) of the Core Package}",
	journal = {\apj},
	year = 2022,
	month = aug,
	volume = {935},
	number = {2},
	eid = {167},
	pages = {167},
	doi = {10.3847/1538-4357/ac7c74},
	archivePrefix = {arXiv},
	eprint = {2206.14220},
	primaryClass = {astro-ph.IM},
	adsurl = {https://ui.adsabs.harvard.edu/abs/2022ApJ...935..167A}
}

@ARTICLE{fu2023,
       author = {{Fu}, Zhen-Sen and {Qi}, Zhao-Xiang and {Liao}, Shi-Long and {Peng}, Xi-Yan and {Yu}, Yong and {Wu}, Qi-Qi and {Shao}, Li and {Xu}, You-Hua},
        title = "{Simulation of CSST's astrometric capability}",
      journal = {Frontiers in Astronomy and Space Sciences},
         year = 2023,
        month = jun,
       volume = {10},
          eid = {1146603},
        pages = {1146603},
          doi = {10.3389/fspas.2023.1146603},
archivePrefix = {arXiv},
       eprint = {2304.02196},
 primaryClass = {astro-ph.IM},
       adsurl = {https://ui.adsabs.harvard.edu/abs/2023FrASS..1046603F}
}

@ARTICLE{euclid ,
       author = {{Euclid Collaboration} and {Moneti}, A. and {McCracken}, H.~J. and {Shuntov}, M. and {Kauffmann}, O.~B. and {Capak}, P. and {Davidzon}, I. and {Ilbert}, O. and {Scarlata}, C. and {Toft}, S. and {Weaver}, J. and {Chary}, R. and {Cuby}, J. and {Faisst}, A.~L. and {Masters}, D.~C. and {McPartland}, C. and {Mobasher}, B. and {Sanders}, D.~B. and {Scaramella}, R. and {Stern}, D. and {Szapudi}, I. and {Teplitz}, H. and {Zalesky}, L. and {Amara}, A. and {Auricchio}, N. and {Bodendorf}, C. and {Bonino}, D. and {Branchini}, E. and {Brau-Nogue}, S. and {Brescia}, M. and {Brinchmann}, J. and {Capobianco}, V. and {Carbone}, C. and {Carretero}, J. and {Castander}, F.~J. and {Castellano}, M. and {Cavuoti}, S. and {Cimatti}, A. and {Cledassou}, R. and {Congedo}, G. and {Conselice}, C.~J. and {Conversi}, L. and {Copin}, Y. and {Corcione}, L. and {Costille}, A. and {Cropper}, M. and {Da Silva}, A. and {Degaudenzi}, H. and {Douspis}, M. and {Dubath}, F. and {Duncan}, C.~A.~J. and {Dupac}, X. and {Dusini}, S. and {Farrens}, S. and {Ferriol}, S. and {Fosalba}, P. and {Frailis}, M. and {Franceschi}, E. and {Fumana}, M. and {Garilli}, B. and {Gillis}, B. and {Giocoli}, C. and {Granett}, B.~R. and {Grazian}, A. and {Grupp}, F. and {Haugan}, S.~V.~H. and {Hoekstra}, H. and {Holmes}, W. and {Hormuth}, F. and {Hudelot}, P. and {Jahnke}, K. and {Kermiche}, S. and {Kiessling}, A. and {Kilbinger}, M. and {Kitching}, T. and {Kohley}, R. and {K{\"u}mmel}, M. and {Kunz}, M. and {Kurki-Suonio}, H. and {Ligori}, S. and {Lilje}, P.~B. and {Lloro}, I. and {Maiorano}, E. and {Mansutti}, O. and {Marggraf}, O. and {Markovic}, K. and {Marulli}, F. and {Massey}, R. and {Maurogordato}, S. and {Meneghetti}, M. and {Merlin}, E. and {Meylan}, G. and {Moresco}, M. and {Moscardini}, L. and {Munari}, E. and {Niemi}, S.~M. and {Padilla}, C. and {Paltani}, S. and {Pasian}, F. and {Pedersen}, K. and {Pires}, S. and {Poncet}, M. and {Popa}, L. and {Pozzetti}, L. and {Raison}, F. and {Rebolo}, R. and {Rhodes}, J. and {Rix}, H. and {Roncarelli}, M. and {Rossetti}, E. and {Saglia}, R. and {Schneider}, P. and {Secroun}, A. and {Seidel}, G. and {Serrano}, S. and {Sirignano}, C. and {Sirri}, G. and {Stanco}, L. and {Tallada-Cresp{\'\i}}, P. and {Taylor}, A.~N. and {Tereno}, I. and {Toledo-Moreo}, R. and {Torradeflot}, F. and {Wang}, Y. and {Welikala}, N. and {Weller}, J. and {Zamorani}, G. and {Zoubian}, J. and {Andreon}, S. and {Bardelli}, S. and {Camera}, S. and {Graci{\'a}-Carpio}, J. and {Medinaceli}, E. and {Mei}, S. and {Polenta}, G. and {Romelli}, E. and {Sureau}, F. and {Tenti}, M. and {Vassallo}, T. and {Zacchei}, A. and {Zucca}, E. and {Baccigalupi}, C. and {Balaguera-Antol{\'\i}nez}, A. and {Bernardeau}, F. and {Biviano}, A. and {Bolzonella}, M. and {Bozzo}, E. and {Burigana}, C. and {Cabanac}, R. and {Cappi}, A. and {Carvalho}, C.~S. and {Casas}, S. and {Castignani}, G. and {Colodro-Conde}, C. and {Coupon}, J. and {Courtois}, H.~M. and {Di Ferdinando}, D. and {Farina}, M. and {Finelli}, F. and {Flose-Reimberg}, P. and {Fotopoulou}, S. and {Galeotta}, S. and {Ganga}, K. and {Garcia-Bellido}, J. and {Gaztanaga}, E. and {Gozaliasl}, G. and {Hook}, I. and {Joachimi}, B. and {Kansal}, V. and {Keihanen}, E. and {Kirkpatrick}, C.~C. and {Lindholm}, V. and {Mainetti}, G. and {Maino}, D. and {Maoli}, R. and {Martinelli}, M. and {Martinet}, N. and {Maturi}, M. and {Metcalf}, R.~B. and {Morgante}, G. and {Morisset}, N. and {Nucita}, A. and {Patrizii}, L. and {Potter}, D. and {Renzi}, A. and {Riccio}, G. and {S{\'a}nchez}, A.~G. and {Sapone}, D. and {Schirmer}, M. and {Schultheis}, M. and {Scottez}, V. and {Sefusatti}, E. and {Teyssier}, R. and {Tubio}, O. and {Tutusaus}, I. and {Valiviita}, J. and {Viel}, M. and {Hildebrandt}, H.},
        title = "{Euclid preparation. XVII. Cosmic Dawn Survey: Spitzer Space Telescope observations of the Euclid deep fields and calibration fields}",
      journal = {\aap},
         year = 2022,
        month = feb,
       volume = {658},
          eid = {A126},
        pages = {A126},
          doi = {10.1051/0004-6361/202142361},
archivePrefix = {arXiv},
       eprint = {2110.13928},
 primaryClass = {astro-ph.GA},
       adsurl = {https://ui.adsabs.harvard.edu/abs/2022A&A...658A.126E}
}

@article{zonca2019,
  doi = {10.21105/joss.01298},
  url = {https://doi.org/10.21105/joss.01298},
  year = {2019},
  month = mar,
  publisher = {The Open Journal},
  volume = {4},
  number = {35},
  pages = {1298},
  author = {Andrea Zonca and Leo Singer and Daniel Lenz and Martin Reinecke and Cyrille Rosset and Eric Hivon and Krzysztof Gorski},
  title = {healpy: equal area pixelization and spherical harmonics transforms for data on the sphere in Python},
  journal = {Journal of Open Source Software}
}

@ARTICLE{gorski2005,
       author = {{G{\'o}rski}, K.~M. and {Hivon}, E. and {Banday}, A.~J. and {Wandelt}, B.~D. and {Hansen}, F.~K. and {Reinecke}, M. and {Bartelmann}, M.},
        title = "{HEALPix: A Framework for High-Resolution Discretization and Fast Analysis of Data Distributed on the Sphere}",
      journal = {\apj},
         year = 2005,
        month = apr,
       volume = {622},
       number = {2},
        pages = {759-771},
          doi = {10.1086/427976},
archivePrefix = {arXiv},
       eprint = {astro-ph/0409513},
 primaryClass = {astro-ph},
       adsurl = {https://ui.adsabs.harvard.edu/abs/2005ApJ...622..759G}
}

@ARTICLE{sirianni2005,
       author = {{Sirianni}, M. and {Jee}, M.~J. and {Ben{\'\i}tez}, N. and {Blakeslee}, J.~P. and {Martel}, A.~R. and {Meurer}, G. and {Clampin}, M. and {De Marchi}, G. and {Ford}, H.~C. and {Gilliland}, R. and {Hartig}, G.~F. and {Illingworth}, G.~D. and {Mack}, J. and {McCann}, W.~J.},
        title = "{The Photometric Performance and Calibration of the Hubble Space Telescope Advanced Camera for Surveys}",
      journal = {\pasp},
         year = 2005,
        month = oct,
       volume = {117},
       number = {836},
        pages = {1049-1112},
          doi = {10.1086/444553},
archivePrefix = {arXiv},
       eprint = {astro-ph/0507614},
 primaryClass = {astro-ph},
       adsurl = {https://ui.adsabs.harvard.edu/abs/2005PASP..117.1049S}
}

@ARTICLE{euclid2022_v1,
       author = {{Euclid Collaboration} and {Scaramella}, R. and {Amiaux}, J. and {Mellier}, Y. and {Burigana}, C. and {Carvalho}, C.~S. and {Cuillandre}, J.-C. and {Da Silva}, A. and {Derosa}, A. and {Dinis}, J. and {Maiorano}, E. and {Maris}, M. and {Tereno}, I. and {Laureijs}, R. and {Boenke}, T. and {Buenadicha}, G. and {Dupac}, X. and {Gaspar Venancio}, L.~M. and {G{\'o}mez-{\'A}lvarez}, P. and {Hoar}, J. and {Lorenzo Alvarez}, J. and {Racca}, G.~D. and {Saavedra-Criado}, G. and {Schwartz}, J. and {Vavrek}, R. and {Schirmer}, M. and {Aussel}, H. and {Azzollini}, R. and {Cardone}, V.~F. and {Cropper}, M. and {Ealet}, A. and {Garilli}, B. and {Gillard}, W. and {Granett}, B.~R. and {Guzzo}, L. and {Hoekstra}, H. and {Jahnke}, K. and {Kitching}, T. and {Maciaszek}, T. and {Meneghetti}, M. and {Miller}, L. and {Nakajima}, R. and {Niemi}, S.~M. and {Pasian}, F. and {Percival}, W.~J. and {Pottinger}, S. and {Sauvage}, M. and {Scodeggio}, M. and {Wachter}, S. and {Zacchei}, A. and {Aghanim}, N. and {Amara}, A. and {Auphan}, T. and {Auricchio}, N. and {Awan}, S. and {Balestra}, A. and {Bender}, R. and {Bodendorf}, C. and {Bonino}, D. and {Branchini}, E. and {Brau-Nogue}, S. and {Brescia}, M. and {Candini}, G.~P. and {Capobianco}, V. and {Carbone}, C. and {Carlberg}, R.~G. and {Carretero}, J. and {Casas}, R. and {Castander}, F.~J. and {Castellano}, M. and {Cavuoti}, S. and {Cimatti}, A. and {Cledassou}, R. and {Congedo}, G. and {Conselice}, C.~J. and {Conversi}, L. and {Copin}, Y. and {Corcione}, L. and {Costille}, A. and {Courbin}, F. and {Degaudenzi}, H. and {Douspis}, M. and {Dubath}, F. and {Duncan}, C.~A.~J. and {Dusini}, S. and {Farrens}, S. and {Ferriol}, S. and {Fosalba}, P. and {Fourmanoit}, N. and {Frailis}, M. and {Franceschi}, E. and {Franzetti}, P. and {Fumana}, M. and {Gillis}, B. and {Giocoli}, C. and {Grazian}, A. and {Grupp}, F. and {Haugan}, S.~V.~H. and {Holmes}, W. and {Hormuth}, F. and {Hudelot}, P. and {Kermiche}, S. and {Kiessling}, A. and {Kilbinger}, M. and {Kohley}, R. and {Kubik}, B. and {K{\"u}mmel}, M. and {Kunz}, M. and {Kurki-Suonio}, H. and {Lahav}, O. and {Ligori}, S. and {Lilje}, P.~B. and {Lloro}, I. and {Mansutti}, O. and {Marggraf}, O. and {Markovic}, K. and {Marulli}, F. and {Massey}, R. and {Maurogordato}, S. and {Melchior}, M. and {Merlin}, E. and {Meylan}, G. and {Mohr}, J.~J. and {Moresco}, M. and {Morin}, B. and {Moscardini}, L. and {Munari}, E. and {Nichol}, R.~C. and {Padilla}, C. and {Paltani}, S. and {Peacock}, J. and {Pedersen}, K. and {Pettorino}, V. and {Pires}, S. and {Poncet}, M. and {Popa}, L. and {Pozzetti}, L. and {Raison}, F. and {Rebolo}, R. and {Rhodes}, J. and {Rix}, H.-W. and {Roncarelli}, M. and {Rossetti}, E. and {Saglia}, R. and {Schneider}, P. and {Schrabback}, T. and {Secroun}, A. and {Seidel}, G. and {Serrano}, S. and {Sirignano}, C. and {Sirri}, G. and {Skottfelt}, J. and {Stanco}, L. and {Starck}, J.~L. and {Tallada-Cresp{\'\i}}, P. and {Tavagnacco}, D. and {Taylor}, A.~N. and {Teplitz}, H.~I. and {Toledo-Moreo}, R. and {Torradeflot}, F. and {Trifoglio}, M. and {Valentijn}, E.~A. and {Valenziano}, L. and {Verdoes Kleijn}, G.~A. and {Wang}, Y. and {Welikala}, N. and {Weller}, J. and {Wetzstein}, M. and {Zamorani}, G. and {Zoubian}, J. and {Andreon}, S. and {Baldi}, M. and {Bardelli}, S. and {Boucaud}, A. and {Camera}, S. and {Di Ferdinando}, D. and {Fabbian}, G. and {Farinelli}, R. and {Galeotta}, S. and {Graci{\'a}-Carpio}, J. and {Maino}, D. and {Medinaceli}, E. and {Mei}, S. and {Neissner}, C. and {Polenta}, G. and {Renzi}, A. and {Romelli}, E. and {Rosset}, C. and {Sureau}, F. and {Tenti}, M. and {Vassallo}, T. and {Zucca}, E. and {Baccigalupi}, C. and {Balaguera-Antol{\'\i}nez}, A. and {Battaglia}, P. and {Biviano}, A. and {Borgani}, S. and {Bozzo}, E. and {Cabanac}, R. and {Cappi}, A.},
        title = "{Euclid preparation. I. The Euclid Wide Survey}",
      journal = {\aap},
         year = 2022,
        month = jun,
       volume = {662},
          eid = {A112},
        pages = {A112},
          doi = {10.1051/0004-6361/202141938},
archivePrefix = {arXiv},
       eprint = {2108.01201},
 primaryClass = {astro-ph.CO},
       adsurl = {https://ui.adsabs.harvard.edu/abs/2022A&A...662A.112E}
}

@ARTICLE{Gong2025,
       author = {{CSST Collaboration} and {Gong}, Yan and {Miao}, Haitao and {Zhan}, Hu and {Li}, Zhao-Yu and {Shangguan}, Jinyi and {Li}, Haining and {Liu}, Chao and {Chen}, Xuefei and {Yuan}, Haibo and {Zhou}, Jilin and {Liu}, Hui-Gen and {Yu}, Cong and {Ji}, Jianghui and {Qi}, Zhaoxiang and {Liu}, Jiacheng and {Dai}, Zigao and {Wang}, Xiaofeng and {Zheng}, Zhenya and {Hao}, Lei and {Dou}, Jiangpei and {Ao}, Yiping and {Lin}, Zhenhui and {Zhang}, Kun and {Wang}, Wei and {Sun}, Guotong and {Li}, Ran and {Li}, Guoliang and {Xu}, Youhua and {Li}, Xinfeng and {Li}, Shengyang and {Wu}, Peng and {Zhang}, Jiuxing and {Wang}, Bo and {Bai}, Jinming and {Cai}, Yi-Fu and {Cai}, Zheng and {Cao}, Jie and {Chan}, Kwan Chuen and {Chang}, Jin and {Chen}, Xiaodian and {Chen}, Xuelei and {Chen}, Yuqin and {Chen}, Yun and {Cui}, Wei and {Dong}, Subo and {Du}, Pu and {Duan}, Wenying and {Fan}, Junhui and {Fan}, LuLu and {Fan}, Zhou and {Fan}, Zuhui and {Fang}, Taotao and {Fu}, Jianning and {Fu}, Liping and {Fu}, Zhensen and {Gao}, Jian and {Gu}, Shenghong and {Gu}, Yidong and {Guo}, Qi and {Han}, Zhanwen and {Hu}, Bin and {Huang}, Zhiqi and {Ho}, Luis C. and {Jiang}, Linhua and {Jiang}, Ning and {Jing}, Yipeng and {Kang}, Xi and {Kong}, Xu and {Li}, Cheng and {Li}, Chengyuan and {Li}, Di and {Li}, Jing and {Li}, Nan and {Li}, Yang A. and {Liao}, Shilong and {Lin}, Weipeng and {Liu}, Fengshan and {Liu}, Jifeng and {Liu}, Xiangkun and {Liu}, Zhuokai and {Mao}, Ruiqing and {Mao}, Shude and {Meng}, Xianmin and {Pang}, Xiaoying and {Peng}, Xiyan and {Peng}, Yingjie and {Shan}, Huanyuan and {Shen}, Juntai and {Shen}, Shiyin and {Shen}, Zhiqiang and {Shi}, Sheng-Cai and {Shi}, Yong and {Tan}, Siyuan and {Tian}, Hao and {Wang}, Jianmin and {Wang}, Jun-Xian and {Wang}, Xin and {Wang}, Yuting and {Wu}, Hong and {Wu}, Jingwen and {Wu}, Xuebing and {Xu}, Chun and {Xue}, Xiang-Xiang and {Xue}, Yongquan and {Yang}, Ji and {Yang}, Xiaohu and {Yao}, Qijun and {Yuan}, Fangting and {Yuan}, Zhen and {Zhang}, Jun and {Zhang}, Pengjie and {Zhang}, Tianmeng and {Zhang}, Wei and {Zhang}, Xin and {Zhao}, Gang and {Zhao}, Gongbo and {Zhong}, Hongen and {Zhong}, Jing and {Zhou}, Liyong and {Zhu}, Wei and {Zu}, Ying},
        title = "{Introduction to the Chinese Space Station Survey Telescope (CSST)}",
      journal = {arXiv e-prints},
         year = 2025,
        month = jul,
          eid = {arXiv:2507.04618},
        pages = {arXiv:2507.04618},
          doi = {10.48550/arXiv.2507.04618},
archivePrefix = {arXiv},
       eprint = {2507.04618},
 primaryClass = {astro-ph.IM},
       adsurl = {https://ui.adsabs.harvard.edu/abs/2025arXiv250704618C}
}

@ARTICLE{Skrutskie2006,
       author = {{Skrutskie}, M.~F. and {Cutri}, R.~M. and {Stiening}, R. and {Weinberg}, M.~D. and {Schneider}, S. and {Carpenter}, J.~M. and {Beichman}, C. and {Capps}, R. and {Chester}, T. and {Elias}, J. and {Huchra}, J. and {Liebert}, J. and {Lonsdale}, C. and {Monet}, D.~G. and {Price}, S. and {Seitzer}, P. and {Jarrett}, T. and {Kirkpatrick}, J.~D. and {Gizis}, J.~E. and {Howard}, E. and {Evans}, T. and {Fowler}, J. and {Fullmer}, L. and {Hurt}, R. and {Light}, R. and {Kopan}, E.~L. and {Marsh}, K.~A. and {McCallon}, H.~L. and {Tam}, R. and {Van Dyk}, S. and {Wheelock}, S.},
        title = "{The Two Micron All Sky Survey (2MASS)}",
      journal = {\aj},
         year = 2006,
        month = feb,
       volume = {131},
       number = {2},
        pages = {1163-1183},
          doi = {10.1086/498708},
       adsurl = {https://ui.adsabs.harvard.edu/abs/2006AJ....131.1163S}
}

@ARTICLE{Zheng2025,
       author = {{Zheng}, Zhen-Ya and {Xu}, Chun and {Liu}, Xiaohua and {Chen}, Yong-He and {Xu}, Fang and {Zhan}, Hu and {Li}, Xinfeng and {Zheng}, Lixin and {Shan}, Huanyuan and {Zhong}, Jing and {Yan}, Zhaojun and {Yuan}, Fang-Ting and {Jiang}, Chunyan and {Peng}, Xiyan and {Chen}, Wei and {Cheng}, Xue and {Chen}, Zhen-Lei and {Zhu}, Shuairu and {Long}, Lin and {Zhang}, Xin and {Gong}, Yan and {Shao}, Li and {Wang}, Wei and {Zhang}, Tianyi and {Ju}, Guohao and {Li}, Chenghao and {Wang}, Wei and {Li}, Zhiyuan and {Wang}, Tao and {Wang}, Junfeng and {Li}, Chengyuan and {Ma}, Bin and {Wang}, Jianguo and {Wang}, Lei and {Liu}, Dezi and {Lin}, Nie and {Li}, Kexin and {Wen}, Xinrong and {Wu}, Maochun and {Lin}, Ruqiu and {Ji}, Xiang},
        title = "{MCI: Multi-Channel Imager on the Chinese Space Station Survey Telescope}",
      journal = {arXiv e-prints},
         year = 2025,
        month = sep,
          eid = {arXiv:2509.14691},
        pages = {arXiv:2509.14691},
          doi = {10.48550/arXiv.2509.14691},
archivePrefix = {arXiv},
       eprint = {2509.14691},
 primaryClass = {astro-ph.IM},
       adsurl = {https://ui.adsabs.harvard.edu/abs/2025arXiv250914691Z}
}

@ARTICLE{Yasuda2007,
       author = {{Yasuda}, Naoki and {Fukugita}, Masataka and {Schneider}, Donald P.},
        title = "{Spatial Variations of Galaxy Number Counts in the Sloan Digital Sky Survey. II. Test of Galactic Extinction in High-Extinction Regions}",
      journal = {\aj},
         year = 2007,
        month = aug,
       volume = {134},
       number = {2},
        pages = {698-705},
          doi = {10.1086/519836},
archivePrefix = {arXiv},
       eprint = {0706.0369},
 primaryClass = {astro-ph},
       adsurl = {https://ui.adsabs.harvard.edu/abs/2007AJ....134..698Y}
}
	
\end{document}